\documentclass[journal]{IEEEtran}
\usepackage{cite}
\usepackage{url}

\usepackage{textcomp}
\usepackage{xcolor}
\usepackage{hyperref}

\usepackage{booktabs}	
\usepackage{diagbox}    
\usepackage{multirow}   

\usepackage{verbatim}	

\usepackage{amsmath,amssymb,amsfonts,graphicx}
\usepackage{epstopdf}

\usepackage{algorithm}
\usepackage{algorithmicx}
\usepackage{algpseudocode}

\newtheorem{definition}{Definition}[section]
\newtheorem{property}{Property}[section]

\def\BibTeX{{\rm B\kern-.05em{\sc i\kern-.025em b}\kern-.08em
		T\kern-.1667em\lower.7ex\hbox{E}\kern-.125emX}}

\usepackage{amssymb}
\usepackage{hyperref}

\usepackage[switch]{lineno}   

\ifCLASSINFOpdf

\else
\fi

\begin{document}

\title{TOPIC: Top-$k$ High-Utility Itemset Discovering} 

\author{Jiahui Chen,~\IEEEmembership{Member,~IEEE,} 
	Shicheng Wan, Wensheng Gan,~\IEEEmembership{Member,~IEEE,}  \\    
	Guoting Chen,	and Hamido Fujita,~\IEEEmembership{Senior Member,~IEEE}
	
\thanks{This work was partially supported by National Natural Science Foundation of China (Grant No. 61902079  and Grant No. 62002136), the Key Areas Research and Development Program of Guangdong Province (Grant No. 2019B010139002),  and the project of Guangzhou Science and Technology (Grant No. 201902020006 and Grant No. 201902020007). (Corresponding author: Wensheng Gan)}
	
	\thanks{Jiahui Chen and Shicheng Wan are with the Department of Computer Sciences, Guangdong University of Technology, Guangzhou 510006, China. (E-mail: csjhchen@gmail.com and scwan1998@gmail.com)}

	\thanks{Wensheng Gan is with the College of Cyber Security, Jinan University, Guangzhou 510632, Guangdong, China; and with Guangdong Artificial Intelligence and Digital Economy Laboratory (Pazhou Lab), Guangzhou 510335, China. (E-mail: wsgan001@gmail.com)}

	\thanks{Guoting Chen is with the School of Science, Harbin Institute of Technology (Shenzhen), Shenzhen, China (E-mail: chenguoting@hit.edu.cn)}

	\thanks{Hamido Fujita is with the Faculty of Software and Information Science,  Iwate Prefectural University, Iwate, Japan (E-mail: HFujita-799@acm.org)}

}


\maketitle


\begin{abstract}	
	Utility-driven itemset mining is widely applied in many real-world scenarios. However, most algorithms do not work for itemsets with negative utilities. Several efficient algorithms for high-utility itemset (HUI) mining with negative utilities have been proposed. These algorithms can find complete HUIs with or without negative utilities. However, the major problem with these algorithms is how to select an appropriate minimum utility (\textit{minUtil}) threshold. To address this issue, some efficient algorithms for extracting top-$k$ HUIs have been proposed, where parameter $k$ is the quantity of HUIs to be discovered. However, all of these algorithms can solve only one part of the above problem. In this paper, we present a method for TOP-$k$ high-utility Itemset disCovering (TOPIC) with positive and negative utility values, which utilizes the advantages of the above algorithms. TOPIC adopts transaction merging and database projection techniques to reduce the database scanning cost, and utilizes \textit{minUtil} threshold raising strategies. It also uses an array-based utility technique, which calculates the utility of itemsets and upper bounds in linear time. We conducted extensive experiments on several real and synthetic datasets, and the results showed that TOPIC outperforms state-of-the-art algorithm in terms of runtime, memory costs, and scalability.
\end{abstract}


%
\IEEEpeerreviewmaketitle

\section{Introduction}

\IEEEPARstart{F}{requent} itemset mining (FIM) \cite{agrawal1994fast,fournier2017survey,gan2017data} has been widely applied in many numerous domains in real-world applications. A classical case of FIM is market basket analysis \cite{han2000mining}. The main target of FIM algorithm is to discover itemsets which are frequently represented; the classic sample is “\textit{Bear} and \textit{Diapers} Theory”. However, simply investigating the frequency of itemsets causes other important factors to be ignored, for example, the quantity of goods that customers purchase or the profit of goods on sale. In FIM \cite{agrawal1994fast, fournier2017survey}, all items/objects are assumed to be the same quantity. For example, if a customer buys a loaf or ten loaves of bread in a store, the two items are regarded as the same quantity in FIM. Moreover, if a customer buys a luxury diamond or a cheap loaf, both goods are also regarded as having the same profit. Obviously, this case does not occur in real life as retailers and managers always focus on finding itemsets that will yield more profits. To address these issues, utility-driven itemset mining, also called high-utility itemset mining (HUIM) \cite{gan2021survey}, has attracted considerable attention.

Utility-driven itemset mining \cite{gan2021survey,liu2005two, tseng2010up, tseng2012efficient, yun2014high, singh2018mining, ahmed2011huc, liu2012mining, lan2014efficient} proposes a new concept called utility (i.e., importance or interest). It considers the quantity of items and their weight value (e.g., unit profit or price). Therefore, it plays a pivotal role in data mining. An itemset is called a high-utility itemset (HUI) if it has a higher utility value than the user-specified minimum utility (\textit{minUtil}) threshold. Utility-driven mining has been widely applied in many practical applications, including user behavior analysis \cite{shie2013mining}, website click-stream analysis \cite{chu2008efficient}, and cross-marketing analysis \cite{yen2007mining}. Compared with FIM algorithms, HUI mining is widely recognized as being more complicated. This is because frequency has the anti-monotonic property, which means that the superset of infrequent itemsets must be infrequent \cite{agrawal1994fast}. In contrast, utility is neither monotonic nor anti-monotonic, and it cannot cut off all non-HUIs and reduce the search space during the mining procedure. To address this limitation, Liu \textit{et al.} \cite{liu2005two} proposed an overestimation method based on the concept of transaction-weighted utilization (\textit{TWU}), which has the downward closure property (anti-monotonic). Subsequently, several HUIM algorithms have adopted the \textit{TWU}-based technique; however, all of them suffer from numerous candidate generations, itemset joining operations, and multiple database scanning. To solve these limitations, scholars have designed tree-based algorithms \cite{tseng2010up, tseng2012efficient, yun2014high, ahmed2009efficient, subramanian2015up} for HUIM. Although tree-based algorithms can mine HUIs without generating many candidates, they need to scan databases more than once and generate numerous conditional sub-trees. All the aforementioned algorithms are categorized as two-phase model algorithms, and their typical feature is that the calculation process of HUIs is divided into two phases: 1) Generate a large number of candidates (or utility-based pattern tree) and 2) Select real high-utility itemsets from a small candidate set.

The above algorithms obviously require a lot of running time and memory because they generate too many candidate itemsets and require multiple database scans. To overcome these limitations, utility-list based algorithms such as HUI-Miner \cite{liu2012mining}, HUP-Miner \cite{krishnamoorthy2015pruning}, HMiner \cite{krishnamoorthy2017hminer}, and FHM \cite{fournier2014fhm} have been proposed. All these algorithms can discover HUIs by constructing an utility-list structure by scanning the database only once and recursively mining HUIs in the memory. Utility-list is a vertical representation of a database, and it stores the key information of itemsets. However, all the previously discussed algorithms may calculate itemsets that do not appear in the database. To solve this limitation, a typical horizontal database projection-based algorithm called EFIM \cite{zida2017efim} was proposed. EFIM absorbs some outstanding ideas from d2HUP \cite{liu2015mining} in that it first proposes a reverse set enumeration tree. Afterward, it modifies the utility upper bounds and uses several pruning strategies to improve the performance. Although much work has been done on HUI mining, some scenarios exist in the real world in which retailers attract customers with promotions or discount tickets, that may cause some items bring negative utility values (i.e., profit) \cite{singh2018high}. In previous algorithms, the utility of HUIs absolutely decreases when considering negative utility items, and the reason is discussed in \cite{chu2009efficient}. The FHN algorithm \cite{lin2016fhn} addresses this problem from the perspective of the utility list \cite{liu2012mining}. EHIN \cite{singh2018mining} calculates HUIs by dividing the items into positive and negative utility items. After obtaining a positive HUI, EHIN tries to add negative utility items and checks whether the extended itemset is still a HUI.

One of the biggest limitations in HUIM domain is the suitability of \textit{minUtil} threshold, while users generally do not know how much they should set. Thus, specifying the \textit{minUtil} threshold is a key task, and it is very challenging because it directly affects the results and performance of a HUIM algorithm. In a study by Wu \textit{et al.} \cite{tseng2015efficient}, they demonstrated that a small change in the \textit{minUtil} threshold will get completely different execution time results. On the one hand, if we set the \textit{minUtil} too low, we will obtain numerous HUIs and excessive amounts of time and memory may be wasted \cite{krishnamoorthy2019mining}. On the other hand, if we set the \textit{minUtil} threshold too high, we will discover few HUIs so that most interesting patterns will be lost. To find an appropriate threshold, users often have to test the algorithm repeatedly, which is a trial-and-error approach. To overcome this drawback, top-$k$ HUI mining was proposed \cite{tseng2015efficient, duong2016efficient, ryang2015top, gan2020tophui}. In top-$k$ HUIM, set $k$ is used instead of threshold, where $k$ indicates that the user deserves the quantity of HUIs. Although top-$k$ pattern mining is practicable, it is more difficult to adopt than \textit{minUtil} threshold for finding complete HUIs. The key point is that top-$k$ pattern mining algorithm needs to store potential top-$k$ patterns in the memory anytime, and it requires \textit{minUtil} to be automatically raised when finding HUIs.

In our literature survey, a state-of-the-art algorithm called TopHUI \cite{gan2020tophui} was used to mine top-$k$ HUIs with or without negative utility values; however, it suffers from mining performance bottleneck such as long execution time and high memory cost. In this paper, we propose an efficient method called \textbf{TOP}-$k$ high-utility \textbf{I}temset dis\textbf{C}overing (TOPIC) with positive and negative utility values to efficiently solve all the above challenges. The novel contributions of TOPIC are that it can effectively discover the exact top-$k$ HUIs with negative utility values in a large database. The main contributions of this study are as follows:

\begin{itemize}
	\item The novel algorithm adopts database-projection and transaction-merging techniques to reduce run time and memory consumption while processing.
	
	\item One of the key challenges is calculating the utility of HUIs without generating numerous candidates. Hence, we adopted an efficient array-based utility-counting technique to obtain the \textit{TWU} and upper bounds of itemsets in linear time.
	
	\item To store the top-$k$ HUIs, we used a priority queue, and employed the \textit{minUtil} threshold-raising strategy for \textit{minUtil} efficiently and automatically increase.
	
	\item Extensive experimental evaluations were conducted on both real and synthetic datasets to evaluate the proposed algorithm. We also compared the performance of TOPIC and TopHUI. The results show that our algorithm is efficient in terms of both run time and memory consumption. Additionally, it is superior to TopHUI for dense datasets.
\end{itemize}

The remainder of this paper is organized as follows: In Section \ref{sec:relatedwork}, related works on traditional utility mining and top-$k$ domains are introduced. In Section \ref{sec:preliminaries}, some basic preliminaries and the problem statement of top-$k$ HUI mining are introduced. Furthermore, a novel TOPIC algorithm is proposed in Section \ref{sec:algorithm}. The experimental results are presented in Section \ref{sec:experiments}, and conclusion and future work are presented in Section \ref{sec:conclusion}.

\section{Related work}
\label{sec:relatedwork}

In this section, we briefly review some studies about high-utility (positive and negative) itemset mining and top-$k$ high-utility itemset mining.

\subsection{High-Utility Itemset Mining}

Since the first HUIM algorithm called Two-Phase \cite{liu2005two} was proposed, researchers have conducted many studies on HUIM algorithms, including UMining \cite{yao2006unified}, IHUP \cite{ahmed2009efficient}, BAHUI \cite{song2014bahui}, and HUI-Miner \cite{liu2012mining}. All HUIM algorithms can be classified into two-phase and single-phase model algorithms. The biggest difference between these two types of algorithms depends on whether they generate numerous candidate itemsets. The Two-Phase algorithm prunes the search space uses \textit{TWU} concept. Similar to FIM, it has the downward closure property, in which the superset of an itemset cannot be a HUI if its \textit{TWU} is less than the \textit{minUtil} threshold. There are two shortcomings of the two-phase model algorithm: 1) it produces numerous candidates; 2) it calculates HUIs by scanning the dataset at least twice. Meanwhile, IHUP \cite{ahmed2009efficient} solves these limitations by constructing a search tree only scan dataset once, and it reduces the number of candidates generated; thus, it performs better than the two-phase model algorithms. However, the upper bound of the IHUP is not sufficiently accurate, and many unpromising supersets are produced because of the overestimated HUIs. Tseng \textit{et al.} \cite{tseng2010up} designed an influential tree-based algorithm called UP-Growth to maintain the information of items to mine HUIs.

In the HUIM domain, the HUI-Miner algorithm \cite{liu2012mining} is a breakthrough work, and it is a single-phase model algorithm. Compared with two-phase model algorithms, the single-phase model algorithm calculates HUIs while generating candidates. HUI-Miner through a new list structure called utility-list, a pair of utility-lists of length $l$-1 intersect to obtain a utility-list of length $l$. It successfully avoids the problem of numerous candidates because of the remaining utility \cite{liu2012mining}. The highlight of HUI-Miner is that it uses utility-list to calculate HUIs in the memory instead of scanning the dataset multiple times. Meanwhile, Zida \textit{et al.} \cite{zida2017efim} proposed the EFIM algorithm, which has a higher accuracy than the \textit{TWU}-based pruning algorithm (two new upper bounds: \textit{local utility} and \textit{revised sub-tree utility}). They designed two new techniques \textit{high-utility database projection} and \textit{high-utility transaction merging} to improve the efficiency of dataset scanning. Additionally, several HUIM algorithms have been extensively studied to extend the effectiveness of HUIM, such as incremental HUIM \cite{gan2018survey}, concise representation-based HUIM \cite{tseng2014efficient,nguyen2019efficient}, top-$k$ HUIM \cite{tseng2015efficient, ryang2015top, krishnamoorthy2019comparative}, HUIM from uncertain data \cite{lin2016efficient,lin2017efficiently}, and so on \cite{mai2017lattice}. More details on utility-oriented pattern mining can be obtained from Gan \textit{et al.} \cite{gan2021survey}. Although much work has already been done in this data mining field, few studies considered negative utility items.

\subsection{High-Utility Itemset Mining with Negative Utilities}

HUINV-Mine \cite{chu2009efficient} is the first level-wise algorithm to explore negative effects in the HUIM domain. It is a two-phase model algorithm, and it adopts a relatively rough method to solve the negative items problem by pruning itemsets that comprise only negative utility items. However, it does not propose an excellent method for dealing with negative utility items. Meanwhile, UP-GNIV \cite{subramanian2015up} is based on set enumeration tree-based concept and does not generate candidates. It is a modification of UP-Growth algorithm \cite{tseng2010up}, and it is used to find interesting patterns that include negative utilities of items. The performance of UP-GNIV is better than that of HUINIV-Mine. Furthermore, Lin \textit{et al.} \cite{lin2016fhn} proposed FHN (modified from FHM \cite{fournier2014fhm}), a utility-list-based mining algorithm, to solve the problems of tree-based algorithms that find long itemsets by recursively searching shorter itemsets. FHN discovers HUIs from a set of transactions in a vertical data format, whereas level-wise and tree-based algorithms use a horizontal data format. The most interesting highlight is the construction of the PNU-list, which is a tuple (\textit{tid}, \textit{putil}, \textit{nutil}, \textit{rputil}) \cite{lin2016fhn}. It also utilizes the EUCS structure \cite{fournier2014fhm}, EUCP strategy \cite{fournier2014fhm}, and LA-Prune strategy \cite{krishnamoorthy2015pruning} to discover HUIs more efficiently. Inspired by the idea of the PNU-list, the EHIN algorithm \cite{singh2018mining} separately lists negative utility items and tries to add negative utility items after calculating positive HUIs. Afterward, it checks whether they are still HUIs. Gan \textit{et al.} \cite{gan2017mining} then proposed a novel algorithm that can discover HUIs with negative utility values from an uncertain dataset. The algorithm constructs a probability utility-list with a positive-and-negative utility (PU$\pm$-list) structure to maintain positive and negative utility items. However, a question that is the biggest limitation of utility-driven pattern mining algorithms arises: how do we appropriately set the minimum utility threshold?

\subsection{Top-$k$ High-Utility Itemset Mining}

As previously mentioned, Wu \textit{et al.} \cite{wu2012mining} demonstrated that a slight difference in the \textit{minUtil} threshold can lead to a significance difference in the number of candidates generated. In top-$k$ HUIM, the parameter $k$ replaces the \textit{minUtil} threshold. Until now, most top-$k$ HUIM algorithms have been based on modifications of previous positive HUIM algorithms \cite{tseng2015efficient,duong2016efficient}. TKU \cite{tseng2015efficient} is an extension of the UP-Growth algorithm with some efficient threshold-raising strategies, and it is also the first top-$k$ HUI mining algorithm. Moreover, REPT \cite{ryang2015top} stores item information through a pre-evaluation matrix in descending order.

Meanwhile, some researchers on top-$k$ HUIM have sought more efficient methods by adopting single-phase model. The TKO \cite{tseng2015efficient} algorithm utilizes the idea of HUI-Miner, and it uses a structure called PE-matrix to increase \textit{minUtil}. Moreover, it adopts the DGU pruning strategy to remove inefficient items during dataset scanning. KHMC \cite{duong2016efficient} is an extension of FHM. By employing three effectively threshold-raising strategies (RIU, CUD, COV), the COV strategy not only prunes the search space in HUI mining but also raises the \textit{minUtil} threshold and optimizes the EUCS structure by using Hash-Map to construct a new EUCST structure. Additionally, KHMC designs a new co-occurrence pruning technique called EUCPT to address the problem of joining operation costs when calculating the utilities of itemsets HUI-Miner does. KHMC performs better than the TKO and REPT algorithms for top-$k$ HUI mining in terms of memory consumption and execution time. Recently, Gan \textit{et al.} \cite{gan2020tophui} proposed a top-$k$ HUIM algorithm called TopHUI, which is the first can work in a transaction database comprising various types of itemsets with positive and negative utilities. It is an extension of THUI \cite{krishnamoorthy2019mining}. However, TopHUI adopts the PNU-list and thus carries on limitations of FHN \cite{lin2016fhn}, as discussed above.

Although there are many studies on HUI mining or top-$k$ HUI mining, few of them focus on how to discover top-$k$ high-utility (positive or negative) itemsets. This study addresses this gap by mining top-$k$ high-utility itemsets with negative utilities. The proposed algorithm adopts novel search-space pruning strategies to effectively find the correct top-$k$ HUIs.

\section{Preliminaries and problem definition}
\label{sec:preliminaries}

In this section, we present some properties and definitions adopted in the TOPIC algorithm. Most of them were introduced by Liu and Qu \cite{liu2012mining}, Zida \textit{et al.} \cite{zida2017efim}, Singh \textit{et al.} \cite{singh2018mining}, and Li \textit{et al.} \cite{li2008isolated}.

\subsection{Basic Definitions}
\begin{definition}(\textbf{Transaction database}) 
	\rm Let $I$ = \{$I_1$, $I_2$, $\dots$, $I_m$\} be a set of distinct items which may be positive or negative. An itemset is defined as a set $X \subseteq I$. $D$ = \{$T_1$, $T_2$, $ \dots$; $ T_n$\} is a transaction database, where each transaction $T_j$ $\in$ $D$ has a unique identifier $j$ called its \textit{TID} (transaction ID) and $n$ is the number of transactions in $D$. Table \ref{table:example} lists the sample database.
\end{definition}

\begin{table}[!htbp]
	\caption{A transaction database}
	\centering
	
	\label{table:example}
	
	\begin{tabular}{cc}
		\toprule
		\textbf{\textit{Tid}} & \textbf{Transaction (item, quantity)} \\
		\midrule
		$T_1$ & $(A, 1) \, (D, 2) \, (E, 1)$ \\
		$T_2$ & $(B, 1) \, (C, 2) \, (D, 6)$ \\
		$T_3$ & $(A, 3) \, (D, 5)          $ \\
		$T_4$ & $(A, 1) \, (E, 1)          $ \\
		$T_5$ & $(B, 1) \, (C, 2) \, (D, 6)$ \\
		$T_6$ & $(B, 1) \, (C, 1) \, (E, 2)$ \\
		\bottomrule
	\end{tabular}
\end{table}

\begin{definition}(\textbf{Internal utility} and \textbf{external utility})
	\rm Let $x$ be an item. $IU(x, T_j)$ is specified as an internal utility (e.g., purchase quantity) of $x$, and $EU(x)$ is specified as an external utility (e.g., unit profit) of $x$. $EU(x)$ represents the relative importance of $x$ to users. Table \ref{table:profits} lists the external utility of each item.
\end{definition}

\begin{table}[th]
	\setlength{\abovecaptionskip}{0.0cm}	
	\setlength{\belowcaptionskip}{0.1cm}
	
	\caption{External utility values}
	\centering
	
	\label{table:profits}
	\setlength{\tabcolsep}{4mm}{
		\begin{tabular}{llllll}
			\toprule
			\textbf{\textit{Item}} 			   & $A$ & $B$  & $C$  & $D$ & $E$ \\
			\midrule
			\textbf{\textit{External utility}} & \$5 & \$-3 & \$-2 & \$6 & \$10 \\
			\bottomrule
	\end{tabular}}
\end{table}

\begin{definition}(\textbf{Utility of an item})
	\rm The utility of item $x$ in transaction $T_j$ is defined as $U(x, T_j)$ = $IU(x, T_j)$ $\times$ $EU(x)$. This indicates how much profit can be generated according to the sale of item $x$ in transaction $T_j$. The utility of item $x$ in $D$ is denoted by $U(x)$ = $\sum_{ x \in T_j \subseteq D}U(X, T_j)$, and it is used to identify the item $x$ that users need. This is a significant HUIM evaluation standard.
\end{definition}

From Table \ref{table:example}, the database contains six transactions ($T_1$, $T_2$, $\dots$, and $T_6$). Transaction $T_2$ consists of items $B$, $C$, and $D$, which have quantities 1, 2, and 6, respectively. Choosing an item $E$ and computing its utility in $D$, $U(E)$ = $U(E, T_1)$ + $U(E, T_4)$ + $U(E, T_6)$ = \$10 + \$10 + \$20 = \$40.

\begin{definition}(\textbf{Utility of an itemset})
	\rm Let an itemset $X$ consist of \{$x_1$, $x_2$, $\ldots$, $x_m$\}. The utility of $X$ in transaction $T_j$ is defined as $U(X, T_j) $ = $\sum_{x_i \in X \land X \subseteq T_j}U(x_i, T_j)$ (1 $\le$ $i$ $\le$ $m$). The utility of an itemset $X$ in a database is defined as $U(X)$ = $\sum_{X \subseteq T_j \in D}U(X, T_j)$ (1 $\le$ $j$ $\le$ $n$).
\end{definition}

For example, if an itemset $X$ = $\{A, D\}$, we obtain $U(\{AD\})$ = $U(\{AD\}, T_1)$ + $U(\{AD\}, T_3)$ = (\$5 $\times$ 1 + \$6 $\times$ 2) + (\$5 $\times$ 3 + \$6 $\times$ 5) = \$62.

\begin{definition}(\textbf{High-utility itemset})
	\rm An itemset $X$ is referred as HUI if $U(X)$ $\ge$ \textit{minUtil} is true. This means that we suppose these itemsets are interesting.
\end{definition}

\begin{definition}(\textbf{Remaining utility})
	\rm Given a transaction $T_j$, all items $x$ that upper than itemset $X$ are defined as $T_j / X$, where $x$ $\in$ $T_j / X $ (the comparison rule here can be replaced by $\succ_T$ and this symbol will be explained further in the next section). The remaining utility of $X$ in transaction $T_j$ is defined as $RU(X, T_j)$ = $\sum_{x_i \in (T_j / X)}U(x_i, T_j)$ \cite{liu2012mining}.
\end{definition}

As mentioned above, many items compose transactions. All these items have utility properties and transactions also have utility properties. Why do we need to calculate transaction utility? In FIM, if an itemset is frequent, all its supersets are also frequent. For example, if we calculate the utility of item $A$ and its supersets $\{A, D\}$, $\{A, E\}$, and $\{A, D, E\}$ in the database as \$25, \$62, \$30 and \$27, we can clearly see that the utility of $\{A, D, E\}$ is less than that of $\{A, D\}$ and $\{A, E\}$ but greater than $\{A\}$. The utility values of itemsets are neither monotonic nor anti-monotonic. Thus, Liu and Qu \cite{liu2012mining} proposed a new concept called \textit{TWU}. Herein, we will introduce the following concepts. 

\begin{definition}(\textbf{Utility of transaction})
	\rm The utility of a transaction $T_j$ is defined as $TU(T_j)$ = $\sum_{x_i \in T_j}U(x_i, T_j)$.
\end{definition}

\begin{table}[th]
	\setlength{\abovecaptionskip}{0.0cm}	
	\setlength{\belowcaptionskip}{0.1cm}
	
	\caption{Utility of transaction}
	\centering
	
	\label{table:TUs}
	\setlength{\tabcolsep}{4mm}{
		\begin{tabular}{lllllll}
			\toprule
			\textbf{\textit{Tid}}      & $T_1$ & $T_2$ & $T_3$ & $T_4$ & $T_5$ & $T_6$ \\
			\midrule
			\textbf{\textit{Utility}}  & \$27  & \$29  & \$45  & \$15  & \$29  & \$15  \\
			\bottomrule
	\end{tabular}}
\end{table}

\begin{definition}(\textbf{Transaction-weighted utilization})
	\rm The transaction-weighted utilization (\textit{TWU}) of itemset $X$ in $D$ is a utility upper bound, which refers to the sum of transaction utilities that contain $X$. It is denoted as \textit{TWU}($X$) and defined as \textit{TWU}($X$) = $ \sum_{X \subseteq T_j \land T_j \subseteq D}TU(T_j)$. 
\end{definition}

We can compute the real utility of a transaction $T_1$ as $TU(T_1)$ = $U(A, T_1)$ + $U(D, T_1)$ + $U(E, T_1)$ = \$27. The computation of the utility of other transactions are given in Table \ref{table:TUs}. By calculating the \textit{TWU} of item $A$, transactions $T_1$, $T_3$, and $T_4$ contain $A$, as given in Table \ref{table:example}. Hence, \textit{TWU}($A$) = $TU(T_1)$ + $TU(T_3)$ + $TU(T_4)$ = \$27 + \$45 + \$15 = \$87. Table \ref{table:TWUs} lists the items and their \textit{TWU} values for the sample transaction database of Table \ref{table:example}.

\begin{table}[th]
	\setlength{\abovecaptionskip}{0.0cm}	
	\setlength{\belowcaptionskip}{0.1cm}
	
	\caption{Transaction weighted utilization}
	\centering
	
	\label{table:TWUs}
	\setlength{\tabcolsep}{4mm}{
		\begin{tabular}{llllll}
			\toprule
			\textbf{\textit{Item}} & $A$  & $B$  & $C$  & $D$   & $E$ \\
			\midrule
			\textbf{\textit{TWU}}  & \$87 & \$73 & \$73 & \$130 & \$57 \\
			\bottomrule
	\end{tabular}}
\end{table}

\begin{property}(\textbf{Transaction-weighted downward closure property})
	If the \textit{TWU} of an itemset $X$ is less than the \textit{minUtil} threshold, then $X$ and all its supersets are low-utility itemsets. This property is usually exploited as a key pruning strategy for mining HUIs. Its proof process is given in \cite{liu2005two}.
\end{property}

\begin{property}(\textbf{\textit{TWU}-based pruning strategy} \cite{liu2005two})
	\rm From the above properties and definitions, it can be inferred that if the \textit{TWU} value of $X$ is less than the user-specified (\textit{minUtil}) threshold (\textit{TWU}($X$) $<$ \textit{minUtil}), then this itemset and its supersets are low-utility itemsets. Subsequently, we can remove them from the search space.
\end{property}

\begin{definition}(\textbf{High transaction-weighted utilization itemset} \cite{liu2005two})
	After pruning the itemset by \textit{TWU}, the remaining itemsets are a set called high transaction-weighted utilization itemsets (HTWUIs), which are potential HUIs. We need to further inspect the HTWUIs to find the true HUIs. If we set \textit{minUtil} to \$70, we can obtain the HTWUIs listed in Table \ref{table:HTWUIs}.
\end{definition}

\begin{table}[htbp]
	\setlength{\abovecaptionskip}{0.0cm}	
	\setlength{\belowcaptionskip}{0.1cm}
	
	\caption{High transaction-weighted utilization itemsets}
	\centering
	
	\label{table:HTWUIs}
	\setlength{\tabcolsep}{4mm}{
		\begin{tabular}{cccc}
			\toprule
			\textbf{\textit{HTWUI}} & \textbf{\textit{Utility}} & \textbf{\textit{HTWUI}} & \textbf{\textit{Utility}} \\
			\midrule
			$\{A\}$ 	& \$70	& $\{C\}$    & \$73	 \\
			$\{A, D\}$  & \$72	& $\{D\}$    & \$130 \\
			$\{B\}$     & \$73  & $\{B, C\}$ & \$73  \\
			\bottomrule
	\end{tabular}}
\end{table}

\subsection{Dealing with Negative Utilities}

\begin{property}(\textbf{ Relationship between positive and negative utility itemsets})
	\rm Given any itemset $X$, the positive utility of $X$ in a transaction or database is defined as \textit{pUtil(X)}, whereas its negative utility is defined as \textit{nUtil(X)}. Therefore, the real utility of an itemset $X$ in a transaction or database is given as \textit{U(X)} = \textit{pUtil}($X$) + \textit{nUtil}($X$). It can be inferred that \textit{pUtil}($X$) $\ge U(X) \ge$ \textit{nUtil}($X$) \cite{lin2016fhn,gan2020utility}.
\end{property}

Most HUIM algorithms \cite{liu2005two, tseng2010up, tseng2012efficient, yun2014high, singh2018mining, liu2012mining, fournier2014fhm} adopt a \textit{TWU}-based pruning strategy. \textit{TWU} not only supports overestimation for mining HUIs but is also used to prune the search space. However, \textit{TWU} cannot be directly applied to items with negative utilities because $\{B\}, \{C\}$ and $\{B, C\}$ are mistaken for \textit{HTWUI} in Table \ref{table:HTWUIs}. To address this error, Chu \textit{et al.} \cite{chu2009efficient} first redefined the utility value of a transaction and \textit{TWU}. We will introduce them in the following.

\begin{definition}(\textbf{Redefined transaction-weighted utilization} \cite{chu2009efficient})
	\rm To avoid the challenges that negative utility items bring, the redefined transaction utility is given as \textit{RTU}$(T_j)$ for a transaction $T_j$, considering only the positive external utility. Thus, \textit{RTU}$(T_j)$ = $\sum_{x \in T_j \land EU(x) > 0}U(x, T_j)$. The redefined transaction-weighted utilization (\textit{RTWU}) of an itemset $X$ is given as \textit{RTWU}$(X)$ = $\sum_{X \subseteq T_j \in D}$\textit{RTU}$(T_j)$.
\end{definition}

\rm If $j$ = 2, \textit{RTU}$(T_2)$ = \textit{TU}$(D)$ = \$6 $\times$ 6 = \$36. In particular, the redefined transaction utility of items with negative utility is set as \$0; hence, we can easily deduce \textit{RTU}$(T_j)$ $\ge$ \textit{TU}$(T_j)$. Assume itemset $X $ = $\{A\}$, transactions $T_1$, $T_3$, and $T_4$ should be considered. Thus, \textit{RTWU}$(A)$ = \textit{RTU}$(T_1)$ + \textit{RTU}$(T_3)$ + \textit{RTU }$(T_4)$ = \$27 + \$45+ \$15 = \$87. Table \ref{table:HUIs} lists the corresponding \textit{RTWU} of the items.

\begin{table}[htbp]
	\setlength{\abovecaptionskip}{0.0cm}	
	\setlength{\belowcaptionskip}{0.1cm}
	
	\caption{Redefined transaction weighted utility}
	\centering
	
	\label{table:RTWU}
	\setlength{\tabcolsep}{4mm}{
		\begin{tabular}{llllll}
			\toprule
			\textbf{\textit{Item}} & $A$ & $B$ & $C$ & $D$  & $E$ \\
			\midrule
			\textbf{\textit{RTWU}} &\$87 &\$92 &\$92 &\$144 &\$62 \\
			\bottomrule
	\end{tabular}}
\end{table}

\begin{property}(\textbf{\textit{RTWU}-based pruning strategy})
	\rm For an itemset $X$, if \textit{RTWU}($X$) $<$ \textit{minUtil}, then $X$ is not a HUI and all supersets of $X$ are low-utility itemsets. The details of the proof can be found in \cite{lin2016fhn}.
\end{property}

\begin{definition}(\textbf{Potential top-$k$ high-utility itemset})
	An itemset is regarded as a potential top-$k$ high-utility itemset (PKHUI) if its estimated utility value (i.e., \textit{TWU}) is higher than the current \textit{minUtil} threshold. In other words, if the \textit{TWU} of this item is higher than the utility of the $k$-th itemset, it may be referred to as a top-$k$ HUI, and the contents of the top-$k$ HUIs will be adjusted.
\end{definition}

\begin{definition}(\textbf{Top-$k$ high-utility itemset})
	\rm An itemset $X$ is called a top-$k$ HUI if there is a list only $k$-1 itemsets which utility values are higher than $U(X)$, and $X$ is the $k$-th highest utility itemset in this list. In particular, $k$ is a user-specified parameter.
\end{definition}

Given an itemset $\alpha$, some items that can be added to $\alpha$ are defined as $E(\alpha)$ = $\{z$ $\mid$ $z \in I \land z \succ x, \forall x \in \alpha\}$ \cite{zida2017efim} (the symbol ``$\succ$" will be explained in the next section). If $k$ = 5, the top five highest utility itemsets containing negative utility items in the sample database are displayed in Table \ref{table:HUIs}, and the final \textit{minUtil} threshold is \$58.

\begin{table}[htbp]
	\setlength{\abovecaptionskip}{0.0cm}	
	\setlength{\belowcaptionskip}{0.1cm}
	
	\caption{top-5 high-utility itemsets}
	\centering
	
	\label{table:HUIs}
	\begin{tabular}{cc}
		\toprule
		\textbf{Itemset} & \textbf{Utility} \\
		\midrule
		$\{D\}$       	 & \$144 \\
		$\{B, D\}$    	 & \$66  \\
		$\{C, D\}$    	 & \$64	 \\
		$\{A, D\}$    	 & \$62  \\
		$\{B, C, D\}$ 	 & \$58  \\
		\bottomrule
	\end{tabular}
\end{table}

\section{The TOPIC Algorithm}
\label{sec:algorithm}

In this section, we present the TOPIC algorithm for mining top-$k$ HUIs with negative utility values. In Subsection \ref{sec:scan_merg_tech}, two efficient database scanning techniques are utilized, namely: database projection and transaction merging. In Subsection \ref{sec:cal_up_bound}, we explain how to calculate the upper bounds (redefined sub-tree utility and redefined local utility) using utility array (UA). In Subsection \ref{sec:riu_strategy}, we propose an efficient and automatic \textit{minUtil} threshold-raising strategy. In Subsection \ref{sec:TOPIC_algo}, we present the pseudo-code of the TOPIC algorithm and describe it in detail.

\subsection{Upper Bounds on Utilities for Pruning Search Space}

\begin{definition}(\textbf{Extension of an itemset} \cite{zida2017efim})
	\rm If an itemset $\alpha$ can be extended into itemset $Y$ = $\alpha \cup \{X\}$, where $X \in 2^{E(\alpha)}$, and $X$ should not be empty. Similarly, if $\alpha$ can be extended with a single itemset $\{z\}$ that contains only one item, $Y$ = $ \alpha \cup \{z\}$, where $z \in E(\alpha)$.
\end{definition}

\begin{definition}(\textbf{Extension of a negative itemset} \cite{singh2018mining})
	\rm Itemset $\alpha$ can be extended to itemset $Y$, $Y$ = $\alpha \cup \{X\}$, where $X$ is a set of items with negative utility.
\end{definition}

The quantity of transactions contain itemset $\alpha \cup \{X\}$ is less or equal than the number of transactions contains itemset $\alpha$. $\alpha$ that extends with positive utility items may be higher or equal to or lower than $U(\alpha)$. However, when $\alpha$ is extended with a negative utility item $\{X\}$, it must be lower than $U(\alpha)$. Furthermore, if $U(\alpha) \ge minUtil$, then we can try to add $\{X\}$ to $\alpha$. If $U(\alpha \cup \{X\})$ is still higher than or equal to \textit{minUtil}, then $\alpha \cup \{X\}$ is a HUI. We can know that if $\alpha$ = $\{A\}$, then in transaction $T_1$, $E(\alpha)$ = $\{D, E\}$ from Table \ref{table:example}. And extensions of $\alpha$ in lexicographical order are $\{A, D\}$, $\{A, E\}$ and $\{A, D, E\}$. \cite{singh2018mining} introduces this rationale and proof.

We set $\succ$ as the total order of items. Our novel algorithm is updated based on the EFIM algorithm, and it explores the search space by using a depth-first search starting from the root (which is an empty set). To make any itemset $\alpha$ become larger, TOPIC recursively appends item $x_{\rm i}$ to $\alpha$ individually through the $\succ$ order. If we only consider the positive items, the $\succ$ order is sorted by increasing \textit{TWU} \cite{liu2012mining, tseng2015efficient}. However, in order to efficiently use the projection technique during the database scanning, each item and original transaction are sorted according to the $\succ$ total order. Moreover, items are sorted by the \textit{RTWU}-ascending order. If the \textit{RTWU} of the items are equal, then the $\succ$ total order follows the lexicographical order. Particularly, negative items always follow positive items in the sorting rule. Afterward, pseudo-projection is performed in each projection; in other words, each projected transaction is represented by an offset pointer on the corresponding original transaction \cite{singh2018mining, zida2017efim}. 

Note that $pUtil(X)$ $\ge$ $U(X)$ $\ge$ $nUtil(X)$. Inspired by previous studies \cite{lin2016fhn,gan2020utility,singh2018mining}, we only take $pUtil(X)$ into account and ignore all items with negative external utility. With this overstatement, then we adopt the following upper-bound concepts in our novel top-$k$ utility mining algorithm.

\begin{definition}(\textbf{Redefined local utility and redefined sub-tree utility})
	\rm The redefined local utility (\textit{RLU}) of item $x$ with respect to an itemset $\alpha$ that may contain both positive and negative utilities is defined as $RLU(\alpha, x)$ = $\sum_{(\alpha \cup \{x\}) \subseteq T_j \land T_j \subseteq D}$[$U(\alpha, T_j)$ + $RU(\alpha, T_j)$], subject to $EU(x) > 0$. The redefined sub-tree utility (\textit{RSU}) of item $x$ with respect to itemset $\alpha$ (the addition of $x$ to $\alpha$ follows the depth-first search of the sub-tree) is defined as follows:  $RSU(\alpha, x)$ = $\sum_{(\alpha \cup x) \subseteq T_j \land T_j \subseteq D}$[$U(\alpha, T_j)$ + $U(x, T_j)$ + $\sum_{i \in T_j \land i \in E(\alpha \cup \{x\})}U(i, T_j)$], subject to $EU(x) > 0$.
\end{definition}

Note that the original concepts of local utility and sub-tree utility are defined in EIFM \cite{zida2017efim}. For example, if $\alpha$ = $\{A\}$, then \textit{RLU}$(A, D)$ = ($U(\{A\}, T_1)$ + \textit{RU}$(\{D\}, T_1)$) + ($U(\{A\}, T_3)$ + \textit{RU}$(\{D\}, T_3)$) = \$15 + \$15 = \$30. If $\alpha$ = $\{A\}$, then \textit{RSU}$(A, D)$ = ($U(A, T_1)$ + $ U(D, T_1)$ + \$0) + ($U(A, T_3) $ + $ U(D, T_3)$ + \$0) = \$17 + \$45 = \$62. Obviously, the negative utility items are not computed here.

\begin{property}(\textbf{Redefined local utility-based overestimation})
	\rm Given an item $x$ and an itemset $\alpha$, where $x$ $\in$ $E(\alpha)$, and $x$ is an extension of $\alpha$, then $x \in X$ ($X$ is a sub-itemset in $E(\alpha)$). Therefore, \textit{RLU}$(\alpha, x) \ge U(X)$ always holds. Furthermore, if \textit{RLU}$(\alpha, x) $ $<$ \textit{minUtil}, then the item $x$ and all extensions of $\alpha$ containing item $x$ have low utility in a sub-tree. Thus, $x$ and its supersets can be pruned to explore all sub-trees of $\alpha$. 
\end{property}

\begin{property}(\textbf{Redefined sub-tree utility-based overestimation})
	\rm Given an item $x$ and an itemset $\alpha$, where $\forall x \in E(\alpha)$, and $x$ can be an extension of $\alpha$, then $x \in X$ ($X$ is a sub-itemset belongs to $E(\alpha)$). Therefore, \textit{RSU}$(\alpha, x) \ge U(X)$ always holds, when dealing with the database which may contain both positive and negative utilities. Furthermore, if \textit{RSU}$(\alpha, x)$ $<$ \textit{minUtil}, then $x$ and all extensions of $\alpha$ that contain $x$ have low utility in the sub-tree. Thus, $x$ and its supersets can be pruned while exploring all sub-trees of $\alpha$. 
\end{property}

The indirectly proof of the above two properties are demonstrated in EFIM \cite{zida2017efim}. It explains why the upper bound \textit{RLU} are tighter than \textit{TWU}. It shows that \textit{RSU} and \textit{RU} are mathematical equivalents. The major difference is their calculation methods are depth-first searching and child itemsets, respectively. Hence, \textit{RSU} cuts off the whole sub-tree of $\alpha$, including nodes $x$, and \textit{RU} prunes only the descendants of $\alpha$. Therefore, we utilized the \textit{RSU} upper bound rather than the \textit{RU} upper bound to prune the search space. Subsequently, we categorized itemset $\alpha$ into \textit{primary}$(\alpha)$ and \textit{secondary}$(\alpha)$.

\begin{definition}(\textbf{Primary and secondary sets} \cite{zida2017efim})
	\rm For an itemset $\alpha$ in a given database, the \textit{primary} items of $\alpha$ are given as \textit{Primary}($\alpha$) = $\{x$ $\mid$ $x$ $\in$ $E(\alpha)$ $\land$ \textit{RSU}($\alpha, x$) $\ge$ \textit{minUtil}\}, and the \textit{secondary} items of $\alpha$ are given as \textit{Secondary}($\alpha$) =$\{x$ $\mid$ $x$ $\in$ $E(\alpha)$ $\land$ \textit{RLU}($\alpha, x$) $\ge$ \textit{minUtil}\}. Because \textit{RLU}$(\alpha, x)$ $\ge$ \textit{RSU}($\alpha, x$), \textit{primary}($\alpha$) $\subseteq$ \textit{secondary}($\alpha$). \textit{Secondary}($\alpha$) indicates items that are extensible, as all items can combine with another distinct item to form an itemset. This means that extendable items and all items in $\alpha$ can be extended by other elements of $E(\alpha)$. In addition, \textit{primary}($\alpha$) indicates items that are searchable, and each item of this set can be an extension element to expand \textit{secondary}($\alpha$) items.
\end{definition}

In particular, the \textit{RSU} upper bound cannot be directly applied in vertical algorithms such as HUI-Miner, FHM, and FHN because once the utility list is established, these algorithms do not need to perform database scanning again

\subsection{Scanning Using Projection and Merging}
\label{sec:scan_merg_tech}

\textbf{Database scanning using projection technique}. This novel algorithm utilizes a database projection technique to reduce the memory consumption and speed up the run time. When an itemset $\alpha$ is considered when depth-first searching and scanning the transactions of database $D$ to calculate the utility of itemsets within the sub-tree of itemset $\alpha$, those items that do not belong to the $\alpha$ extension are pruned. Database without these items (which is pruned) is called projected database \cite{fournier2014fhm, zida2017efim, lin2016fhn}.

\begin{definition}(\textbf{Projected transaction and projected database} \cite{zida2017efim})
	\rm For an itemset $\alpha$, the projected transaction $T_j$ is defined as $\alpha$-$T_j$ = $\{x | x \in T_j \land x \in E(\alpha)\}$. The projected database $D$ is defined as $\alpha$-$D$ = $\{\alpha$-$T_j | T_j \in D \land \alpha$-$T_j \not=$ $\emptyset\}$. As given in Table \ref{table:example}, if an itemset $\alpha$ = $\{A\}$, then the projected database $\alpha$-$T_1$ = $\{D, E\}$, $\alpha$-$T_3$ = $\{D\}$, and $\alpha$-$T_4$ = $\{E\}$. $\alpha$-$D$ contains these transactions.
\end{definition}

\textbf{Database scanning using merging technique}. Our novel algorithm also utilizes the transaction-merging technique to reduce the database scanning cost. After the database is projected, some identical transactions (which may contain the same items but do not have the same internal utility values) or empty transactions may exist. Merging technique is used to replace these identical transactions with a single transaction \cite{singh2018mining, zida2017efim}. If $T_{\rm i}$ is identical to $T_j$, it represents two transactions containing the same items. However, they may not have the same internal utility (purchase quantity) for each item.

\begin{definition}(\textbf{Transaction merging} \cite{krishnamoorthy2015pruning})
	\rm In a database $D$, several identical transactions such as \{$T_{j_1}$, $T_{j_2}$, $\dots$, $T_{j_n}$\} are replaced by a new transaction $T_M$ = $T_{j_1}$ = $T_{j_2}$ = $\dots$ = $T_{j_n}$. The quantity of each item $x$ in these identical transactions is $IU(x, T_M)$ = $\sum_{1 \le i \le n} IU(x, T_{j_i})$.
\end{definition}

\rm For instance, we can observe from Table \ref{table:example} that transactions $T_2$ and $T_5$ are identical. After merging the transactions, a new transaction $T_{25}$ is obtained, where $IU(B, T_{25})$ = 2, $IU(C, T_{25})$ = 4, and $IU(D, T_{25})$ = 12.

\begin{definition}(\textbf{Projected transaction merging} \cite{zida2017efim})
	\rm If there are several identical projected transactions such as \{$T_{j_1}$, $T_{j_2}$, $\dots$, $T_{j_n}$\}, they are replaced by a new transaction $T_M$ = $T_{j_1}$ = $T_{j_2}$ = $\dots$ = $T_{j_n}$ in database $\alpha$-$D$. The internal utility of each item $x \in T_M$ is defined as $IU(x, T_M)$ = $\sum_{1 \le i \le n}IU(x, T_{j_i})$.
\end{definition}

\rm For example, if an itemset $\alpha$ = $\{A\}$, then the projected database $\alpha$-$D$ contains transactions $\alpha$-$T_1$ = $\{D, \, E\}$, $\alpha$-$T_2$ = $\emptyset$, $\alpha$-$T_3$ = $\{D\}$, $\alpha$-$T_4$ = $\{E\}$, $\alpha$-$T_5$ = $\emptyset$, and $\alpha$-$T_6$ = $\emptyset$. Thus, transactions $\alpha$-$T_2$, $\alpha$-$T_5$, and $\alpha$-$T_6$ can be replaced by a new transaction $T_{256}$ = $\emptyset$.

When identifying identical transactions, a naive method is used to compare each transaction, which is inefficient. To make the transaction merging technique more efficient, we adopt a new total order $\succ_T$ on the transactions in the database before merging \cite{singh2018mining, zida2017efim}.

\begin{definition}(\textbf{Total order on transactions} \cite{zida2017efim})
	\rm The $\succ_T$ order is defined as the lexicographical order when reading all transactions from back to front. Further details about $\succ_T$ follow the EFIM algorithm \cite{zida2017efim}.
\end{definition}

\rm If there are three transactions $T_x$ = $\{a, b, c\}$, $T_y$ = $\{a, b, e\}$, and $T_z$ = $\{a, b\}$, then $T_y $ $\succ_T $ $T_x$ $\succ_T $ $ T_z$.

\begin{property}(\textbf{Transaction order in $\succ_T$-sorted database} \cite{zida2017efim})
	\rm If there is an itemset $\alpha$ and $\succ_T$-sorted database $D$, identical transactions appear consecutively in the projected database $\alpha$-$D$.
\end{property}

\begin{IEEEproof} First, while reading the transactions backward, all of them are sorted in lexicographical order. Second, projections always prune the lowest items of a transaction in lexicographical order. For more details and analysis, refer to Ref. \cite{zida2017efim}.
\end{IEEEproof}

\subsection{Calculation of Upper Bounds using Utility Array}
\label{sec:cal_up_bound}

Novel upper bounds are vital for pruning the search space. After searching the utility itemset mining literature, we utilize an array-based structure called \textit{UA}.

\begin{definition}(\textbf{Utility array} \cite{zida2017efim})
	\rm In a database $D$, there is a set of items $I$. The array element for an item $x$ in the array is given as \textit{UA}$[x]$. Each element stores the utility value of the item $x$, and \textit{UA} has a length of $|I|$.
\end{definition}

\rm \textbf{Calculate \textit{RLU($\alpha$)} using UA}. First, \textit{UA} is initialized by filling all the elements with 0. Second, \textit{UA}$[x]$ = \textit{UA}$[x]$ + $U(\alpha, T_j)$ + \textit{RU}$(\alpha, T_j)$, where $x \in T_j$ $\cap$ $E(\alpha) \land \forall T_j \subseteq D$. After database scanning, $\forall x \in E(\alpha)$, \textit{UA}$[x]$ = \textit{RLU}$(\alpha, x)$, which gives the local utility of all positive itemsets.

\rm \textbf{Calculate \textit{RSU($\alpha$)} using UA}. First, \textit{UA} is initialized by filling all the elements with 0. Second, \textit{UA}$[x]$ = \textit{UA}$[x]$ + $\sum_{I \in T_j \land I \in E(\alpha \cup x)}U(I, T_j)$ + $U(\alpha, T_j)$ + $U(x, T_j)$, where item $x \in T_j$ $\cap $ $E(\alpha)$ $\land$ $\forall T_j$ $\subseteq$ $D$. After database scanning, $\forall x \in E(\alpha)$, \textit{UA}$[x]$ = \textit{RSU}$(\alpha, x)$. 

According to the UA technique, we can obtain the upper bounds of utility in linear time. For more details and comparisons, refer to \cite{singh2018mining}.

\subsection{Threshold Raising Strategy}
\label{sec:riu_strategy}

A key method is to automatically increase the (\textit{minUtil}) threshold, and our new algorithm sets the \textit{minUtil} threshold value to 1 at the beginning. The TopHUI algorithm \cite{gan2020tophui} proposes that the threshold should be raised based on the RTU (raising threshold based on transaction utilities) strategy. The REPT \cite{ryang2015top} introduces a real item utilities (RIU) threshold raising strategy. TOPIC also utilizes it to increase the \textit{minUtil}. Other \textit{minUtil} raising strategies and their detailed discussion are given in \cite{gan2020tophui}.

\begin{algorithm}[htb]
	\caption{RIU strategy}
	\label{alg_riu_strategy}
	\textbf{Input:} \textit{top-$k$ list}: a list of utility values for all items, $k$: the desired number of HUIs. \\
	\textbf{Output:} \textit{minUtil}.
	\begin{algorithmic}[1]
		\State sort top-$k$ list by descending order;
		\If{$\mid$ top-$k$ list $\mid$ $\ge$ $k$}
		\State set the $k$-{th} highest value as a new current \textit{minUtil};
		\EndIf
		\State \Return \textit{minUtil}
	\end{algorithmic}
\end{algorithm}

Here, we provide a brief introduction of \textbf{Algorithm \ref{alg_riu_strategy}}. After calculating $\sum_{T_{\rm j} \in D}U(x, T_{\rm j})$ for all the items, it is added to the top-$k$ list as an input parameter. The subscript $k$ indicates that the user specifies the number of HUIs they need. Afterward, all elements in the top-$k$ list are sorted in descending order. This operation will help us obtain the $k$ highest utility for convenience (Line 1). If the length of the top-$k$ list is higher than $k$, then the current \textit{minUtil} is raised to the $k$-{th} highest value (Lines 2–4). Finally, we obtain a new \textit{minUtil} as the output (Line 5).

\subsection{The TOPIC Algorithm}
\label{sec:TOPIC_algo}

The proposed algorithm TOPIC \textbf{(Algorithm \ref{alg_TOPIC})} adopts some new techniques mentioned in the previous sections. It mainly takes a transaction database and a user-specific parameter $k$ as input parameters and returns the top-$k$ HUIs. In Lines 1–4 of the algorithm, the empty itemsets are separately initialized as $\alpha$. $\rho$ stores a set of positive and negative utility items in the database as $\eta$, and the minimum utility threshold value is 1. In Line 5, a $k$ priority queue is created to maintain a ``candidate” \textit{minUtil} to raise the \textit{minUtil}. In Line 6, the real utility values of all items $z \in I$ is computed and a list \textit{RIU} is used to store these values. Subsequently, the threshold-raising utility function is called to increase the current \textit{minUtil} threshold (Line 7). Afterward, the RLU of each item is calculated using an array (Line 8), and it prepares to select items that can be expanded. Items whose RLUs are higher than the current \textit{minUtil} are then selected to form the \textit{secondary} set (Line 9), and the \textit{secondary} items are sorted in ascending order of \textit{RTWU} (Line 10). Negative utility items are always followed by positive utility items in the algorithm. In Line 11, all low-utility items are removed based on database scanning (\textit{RTWU}-based pruning strategy). Afterward, empty transactions are deleted (Line 12) because there may be some transactions that have only items that are already removed in Line 11. Thereafter, the remaining transactions are sorted by $\succ_{\rm T}$ using lexicographical order in Line 13. Transaction merging is performed in Line 14, and in Line 15, the remaining transactions are scanned again and a UA is used to calculate \textit{RSU}($\alpha, z$), where items $z$ $\in$ \textit{secondary}($\alpha$). A new set of \textit{primary}($\alpha$) items is then obtained (Line 16), which will help to prune the search sub-tree. In Line 17, the negative utility items are stored in the global variable because it needs to try to add these items in the HUI to consider whether it would still be a HUI. The \textbf{search\_P} procedure is called in Line 18 starting with itemset $\alpha$ in the depth-first search. Finally, the top-$k$ high-utility itemsets are returned.

\begin{algorithm}[!h]
	\caption{Proposed TOPIC algorithm}
	\label{alg_TOPIC}
	\begin{algorithmic}[1]
		\Require $D$: a database, $k$: the desired number of HUIs.
		\Ensure Top-$k$ HUIs with negative utility items.
		
		\State initialize $\alpha \gets \emptyset \,$;
		\State initialize $\rho \gets$ a set of positive utility items;
		\State initialize $\eta \gets$ a set of negative utility items;
		\State initialize \textit{minUtil} $\gets 1$
		\State create a priority queue of size $k$\;
		\State compute real utility of all items $z \in I$, and store values into list \textit{RIU};
		\State call \textbf{RIU}(\textit{RIU, k}) to raise the \textit{minUtil};
		\State scan all transactions, using utility-array to calculate  \textit{RLU}($\alpha, z$) of all items $z \in \rho$;
		\State \textit{Secondary}($\alpha$) = $\{z | z \in \rho $ $\land$ \textit{RLU}$(\alpha, z) \ge$ \textit{minUtil}\};
		\State sorted \textit{Secondary}($\alpha$) by using the total order $\succ$ of  \textit{RTWU} increasing values;
		\State scan $D$, remove low utility items $x \not\in $ \textit{Secondary}($\alpha$) from transactions;
		\State remove all empty transactions;
		\State sort all remaining transactions according to the $\succ_T$ using lexicographical order;
		\State assign offset to each transaction in $D$;
		\State scan all remaining transactions in $D$, using utility-array to calculate \textit{RSU}($\alpha, z$) for all items $z \in$ \textit{Secondary}($\alpha$); 
		\State calculate \textit{Primary}$(\alpha)$ = $\{z | z \in $\textit{Secondary}$(\alpha)$ $\land$ \textit{RSU}($\alpha, z)$ $\ge $ \textit{minUtil}\};
		\State store the negative items in global variate;
		\State call \textbf{search\_P}($\eta$, $\alpha$, $D$, \textit{Primary}($\alpha$), \textit{Secondary}($\alpha$), \textit{minUtil}, $k$-\textit{patterns});
		\State\Return top-$k$ HUIs
	\end{algorithmic}
\end{algorithm}

The \textbf{Algorithm \ref{alg_search_P}} has seven input parameters: $\alpha$ is the current itemset prepared to be extended (it is initialized as an empty set), $\eta$ denotes a set of negative utility items, $\alpha$-$D$ is the current projected database (it is initially an original database), the \textit{primary} set contains primary items of itemset $\alpha$, the \textit{secondary} set contains secondary items of itemset $\alpha$, \textit{minUtil} represents the raised minimum utility threshold, and \textit{k patterns} is a priority queue of $k$ items. This algorithm recursively calls itself to extend each positive item of $\alpha$ to constantly find extensions of $\alpha$. Line 2 starts traversing each item $z \in \ $\textit{primary}($\alpha$), and these are regarded as extensible items. In Line 3, each item $z$ is combined with $\alpha$ to form a new itemset $\beta$. Based on the scanned database $\alpha$-$D$, the utility of itemset $\beta$ is calculated and a new merging and projection database $\beta$-$D$ is created. Lines 4-10 show that if the utility value of $\beta$ is higher than or equal to the current \textit{minUtil}, $\beta$ will be recognized as a HUI and added to the \textit{top-k list}. Moreover, if the size of the \textit{top-k list} is larger than $k$, it indicates that top-$k$ HUIs already exist. In this case, the $k$-{th} HUI will be removed and the current \textit{minUtil} will be changed. Lines 11-13 show that if the utility of $\beta$ is also higher (not equal) than the changed \textit{minUtil}, we will try to add negative utility items to verify whether it will still be HUIs, because after itemset $\beta$ extended some negative utility items, its utility may be still higher than the current \textit{minUtil}. Moreover, similar to \textbf{Algorithm \ref{alg_TOPIC}}, \textit{RLU} and \textit{RSU} of itemset $\beta$ are computed, where items $z$ $\in$ \textit{secondary}($\alpha$) (Line 14). In Lines 15 and 16, the \textit{primary} and \textit{secondary} sets of $\beta$ are separately calculated. Finally, the algorithm is repeatedly executed with an extension of $\beta$ using a depth-first search (Line 17) until it satisfies the threshold.

\begin{algorithm}[!h]
	\caption{The \textit{search\_P} procedure}
	\label{alg_search_P}
	\textbf{Input:} $\alpha$: the current itemset, $\eta$: a set of negative items, $\alpha$-$D$: the current projected database, \textit{Primary}($\alpha$): the \textit{Primary} items of $\alpha$, \textit{Secondary}($\alpha$): the \textit{Secondary} items of $\alpha$, \textit{minUtil}: a raised minimum utility threshold, and \textit{top-k list}: a priority queue of $k$ items. \\
	\textbf{Output:} a set of top-$k$ HUIs that are extensions of $\alpha$ with positive utility items.
	
	\begin{algorithmic}[1]	
		\For{each item $z \in \ $\textit{Primary}($\alpha$)}
		\State $\beta = \alpha \cup \{z\}$;
		\State scan $\alpha$-$D$, calculate $U(\beta)$, and create $\beta$-$D$;
		\If{$U$($\beta$) $\ge$ \textit{minUtil}}
		\State add $\beta$ into \textit{top-k list};
		\If{$\mid$\textit{top-k list}$\mid$ $>$ \textit{k}}
		\State pop the $k$-{th} values in \textit{top-k list};
		\State raise current \textit{minUtil} with the $k$-{th} value;
		\EndIf
		\EndIf
		\If{$U$($\beta$) $>$ \textit{minUtil}}
		\State call \textbf{search\_N}($\eta, \, \beta, \, \beta-D, \, $\textit{minUtil}).
		\EndIf
		\State scan $\beta$-$D$, calculate \textit{RSU}($\beta, z$), and \textit{RLU}($\beta, z$) where items $z \in $\textit{Secondary}($\alpha$), using two \textit{UAs};
		\State obtain \textit{Primary}$(\beta)$ = $\{z \in $ \textit{Secondary}$(\alpha$) $\mid$ \textit{RSU}($\beta$, $z) \ge$  \textit{minUtil}\};
		\State obtain \textit{Secondary}($\beta$) = $\{z \in $ \textit{Secondary}($\alpha$) $\mid$ \textit{RLU}($\beta$, $z)$ $\ge$ \textit{minUtil}\};
		\State call \textbf{search\_P}($\eta$, $\beta$, $\beta$-$D$, \textit{Primary}$(\beta)$, \textit{Secondary}$(\beta)$, \textit{minUtil});
		\EndFor	
	\end{algorithmic}
\end{algorithm}

\begin{algorithm}[!h]
	
	\caption{The \textit{search\_N} procedure}
	\label{alg_search_N}
	\textbf{Input:} $\eta$: a set of negative items, $\alpha$: the current itemset, $\alpha$-$D$: the current projected database, \textit{Primary}($\alpha$): the \textit{Primary} items of $\alpha$, \textit{Secondary}($\alpha$): the \textit{Secondary} items of $\alpha$, \textit{minUtil}: a raised minimum utility threshold, and \textit{top-k list}: a priority queue of $k$ items.\\
	\textbf{Output:} The set of top-$k$ HUIs that are extensions of $\alpha$ with negative utility items.
	
	\begin{algorithmic}[1]
		\For{each item $z \in \eta$}
		\State $\beta$ = $\alpha \cup \{z\}$;
		\State scan $\alpha$-$D$, calculate $U(\beta)$, and create $\beta$-$D$;
		\If{$U$($\beta$) $\ge$ \textit{minUtil}}
		\State add $\beta$ into \textit{top-k list};
		\If{$\mid$\textit{top-k list}$\mid$ $>$ \textit{k}}
		\State pop the $k$-{th} values in \textit{top-k list};
		\State raise current \textit{minUtil} with the $k$-{th} value;
		\EndIf
		\EndIf
		\State calculate \textit{RSU}$(\beta, z)$ for all items $z \in \eta$ by scanning itemset $\beta$-$D$ once, using the negative utility-array; 
		\State \textit{Primary}$(\beta)$ = $\{z \in \eta \mid$ \textit{RSU}$(\beta, z)$ $\ge$ \textit{minUtil}\};
		\State call \textbf{search\_N}(\textit{Primary}$(\beta), \beta, \beta$-$D$, \textit{minUtil});
		\EndFor
	\end{algorithmic}
\end{algorithm}

The \textbf{Algorithm \ref{alg_search_N}} is called when the utility of items/itemsets is greater than \textit{minUtil} (not equal). Many of the steps are the same as in \textbf{Algorithm \ref{alg_search_P}}. The main difference is that positive or negative utility items are extended to single items. Each item $z$ combines with $\alpha$ to form a new itemset $\beta$, where each item $z$ $\in$ $\eta$ (Line 2). In Line 3, the database $\alpha$-$D$ is scanned, the utility of extended itemset $\beta$ is computed, and a new projected database $\beta$-$D$ is constructed. Moreover, transaction merging technique is adopted in the database $\beta$-$D$ construction process. Lines 4-10 consider whether the threshold is raised. Subsequently, the \textit{RLU} and \textit{RSU} are calculated again for all negative utility items and a new \textit{primary} set is obtained in Lines 11 and 12. Thereafter, the algorithm recursively calls itself until it does not discover all extensions with negative utility items that satisfy the threshold of \textit{minUtil} (Line 13).

\section{Performance Evaluation}
\label{sec:experiments}

In this section, we conducted several experiments to demonstrate the effectiveness and efficiency of the proposed TOPIC algorithm. We conducted the experiment on a computer with a 3.0 GHz Intel Core Processor with 16 GB main memory running on Windows 10 Home Edition (64-bit operating system). We used Java language to implement all the algorithms and compared the performance of TOPIC with TopHUI \cite{gan2020tophui}. Most of the existing top-$k$ HUIM algorithms do not consider the common real case with negative utility values except TopHUI. To the best of our knowledge, TopHUI is the most efficient algorithm for mining top-$k$ HUIs with negative utilities.

\subsection{Data Description and Experimental Setup}

To analyze the proposed algorithm in different situations, we evaluated its performance on several benchmark datasets. All datasets were downloaded from the SPMF data mining library \cite{fournier2016spmf}. Table \ref{table:Datasets} summarizes the detailed characteristics of all the datasets. The \textit{Mushroom} and \textit{Chess} datasets are highly dense in nature. \textit{Chess} is a dense dataset with long transactions and few items. Although \textit{Mushroom} is also a dense dataset, it has moderately long transactions. Additionally, \textit{Retail} is a sparse dataset with large items in each transaction. \textit{Accidents} is a dense dataset and has the highest number of transactions, with each transaction having many items. \textit{T10I4D100K} and \textit{T40I10D100K} are both sparse datasets. \textit{BMSPOS} is a dense dataset that was used to test the scalability of the proposed algorithm. All the experimental results of these benchmark datasets are separately presented in the following sections. The runtime consumption, memory cost, number of visited candidate itemsets, and scalability are described in subsections \ref{sec:runtime}, \ref{sec:memory}, \ref{sec:candidates}, and \ref{sec:scalability}, respectively.

\begin{table}[!h]
	\setlength{\abovecaptionskip}{0.0cm}	
	\setlength{\belowcaptionskip}{0.1cm}
	
	\caption{Dataset characteristics}
	\centering
	
	\label{table:Datasets}
	\setlength{\tabcolsep}{3.5mm}{
		\begin{tabular}{lllll}
			\toprule
			\textbf{Dataset} & \textbf{\#Trans} & \textbf{\#Items} & \textbf{\#AvgLen} & \textbf{\#Type}  \\
			\midrule
			Mushroom    & 8,142    & 119    & 23.0  & Dense  \\
			Chess       & 3,196    & 75     & 37.0  & Dense  \\
			Accidents   & 340,183  & 468    & 33.8  & Dense  \\
			T40I10D100K & 100,000  & 942    & 39.6  & Dense  \\
			T10I4D100K  & 100,000  & 870    & 10.1  & Sparse \\
			Retail      & 88,162   & 16,470 & 10.3  & Sparse \\
			BMSPOS		& 515,366  & 1,656  & 6.51  & Sparse \\ 
			\bottomrule
	\end{tabular}}
\end{table}

We tested both TopHUI and TOPIC on all datasets by increasing $k$. The \textit{minUtil} was initialized as 1, and we implemented the TopHUI according to the descriptions provided in this paper. We implemented four versions of TOPIC: one with a transaction merging strategy, one with a sub-tree pruning strategy, one implemented with both merging and sub-tree pruning strategy, and the last one is the base version without these two strategies. These versions were referred to as TOPIC$_{merge}$, TOPIC$_{sub-tree}$, TOPIC, and TOPIC$_{none}$, respectively. All the algorithms were used for the experimental evaluation of the proposed top-$k$ HUI mining method.

\begin{figure*}[!h]
	\centering 
	\scalebox{0.46}{\includegraphics[trim={7.5cm 0 7cm 0}, width=32cm, height=19cm]{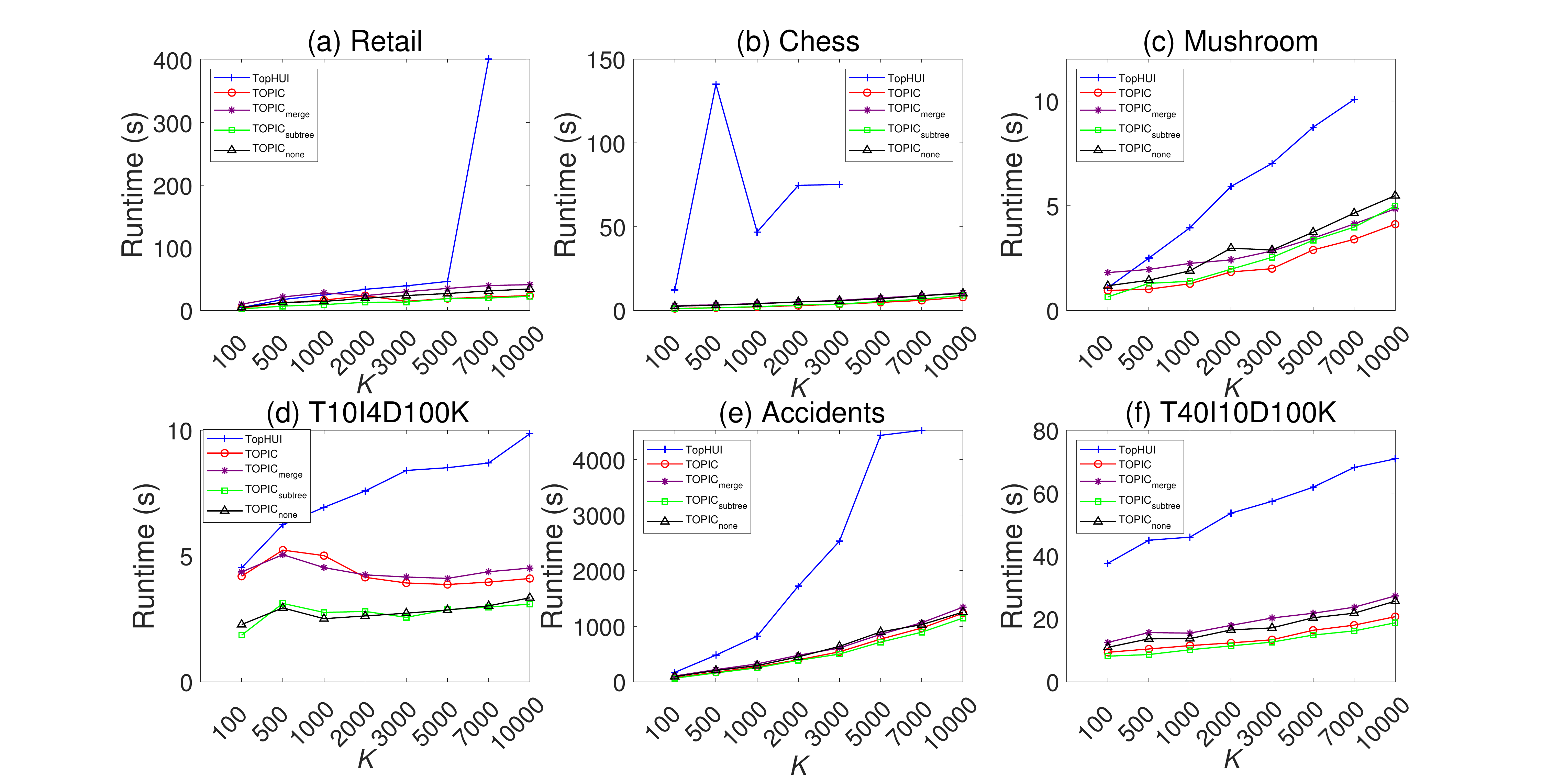}}
	\caption{Runtime cost under parameter (\textit{K}).}
	\label{fig:Runtime}	
\end{figure*}

\begin{figure*}[!h]
	\centering
	\scalebox{0.46}{\includegraphics[trim={7cm 0 7cm 0}, width=32cm, height=19cm]{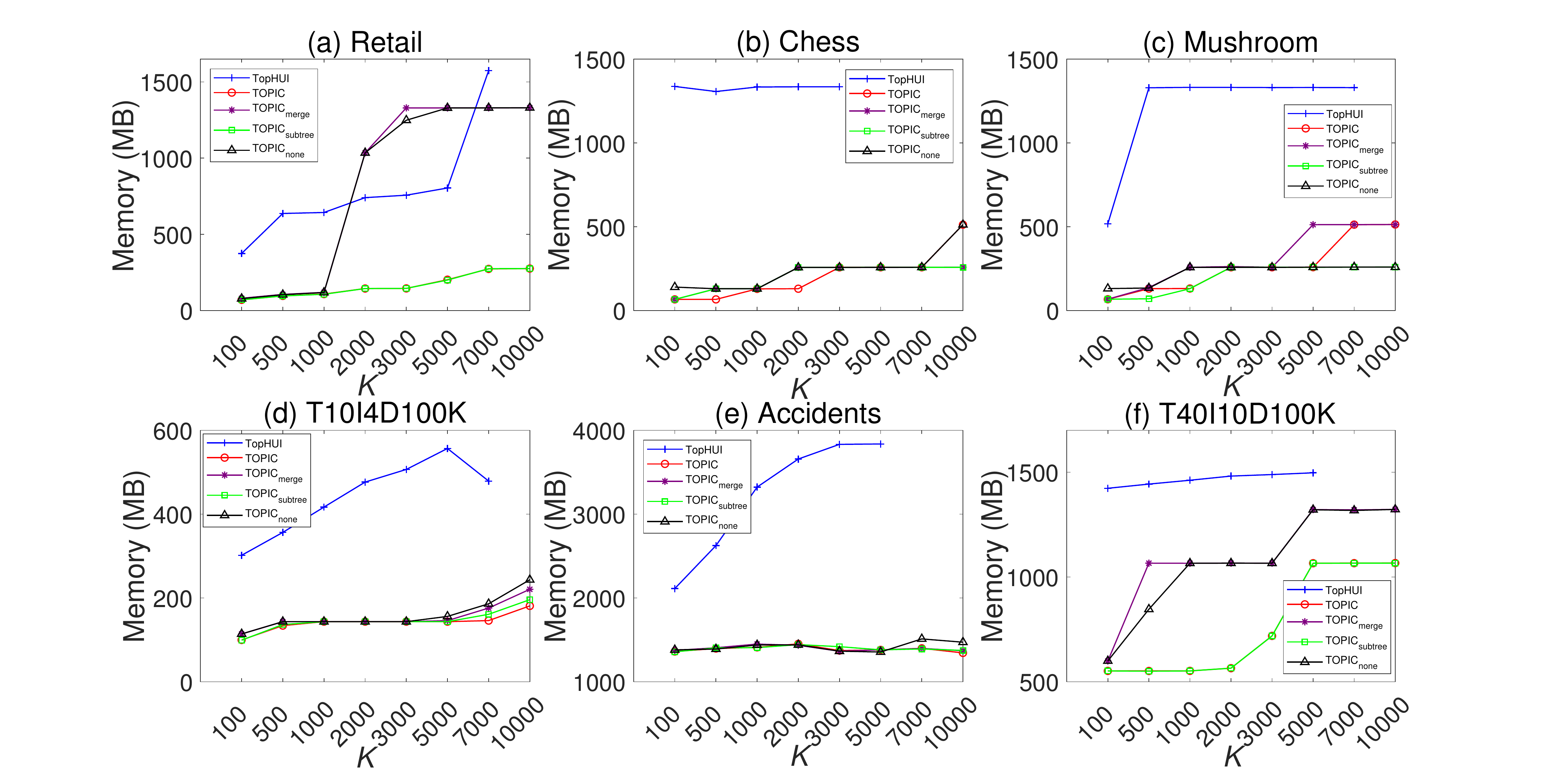}}
	\caption{Memory cost under parameter (\textit{K}).}
	\label{fig:Memory}	
\end{figure*}

\begin{figure*}[h]
	\centering
	\scalebox{0.46}{\includegraphics[trim={7.5cm 0 6.8cm 0}]{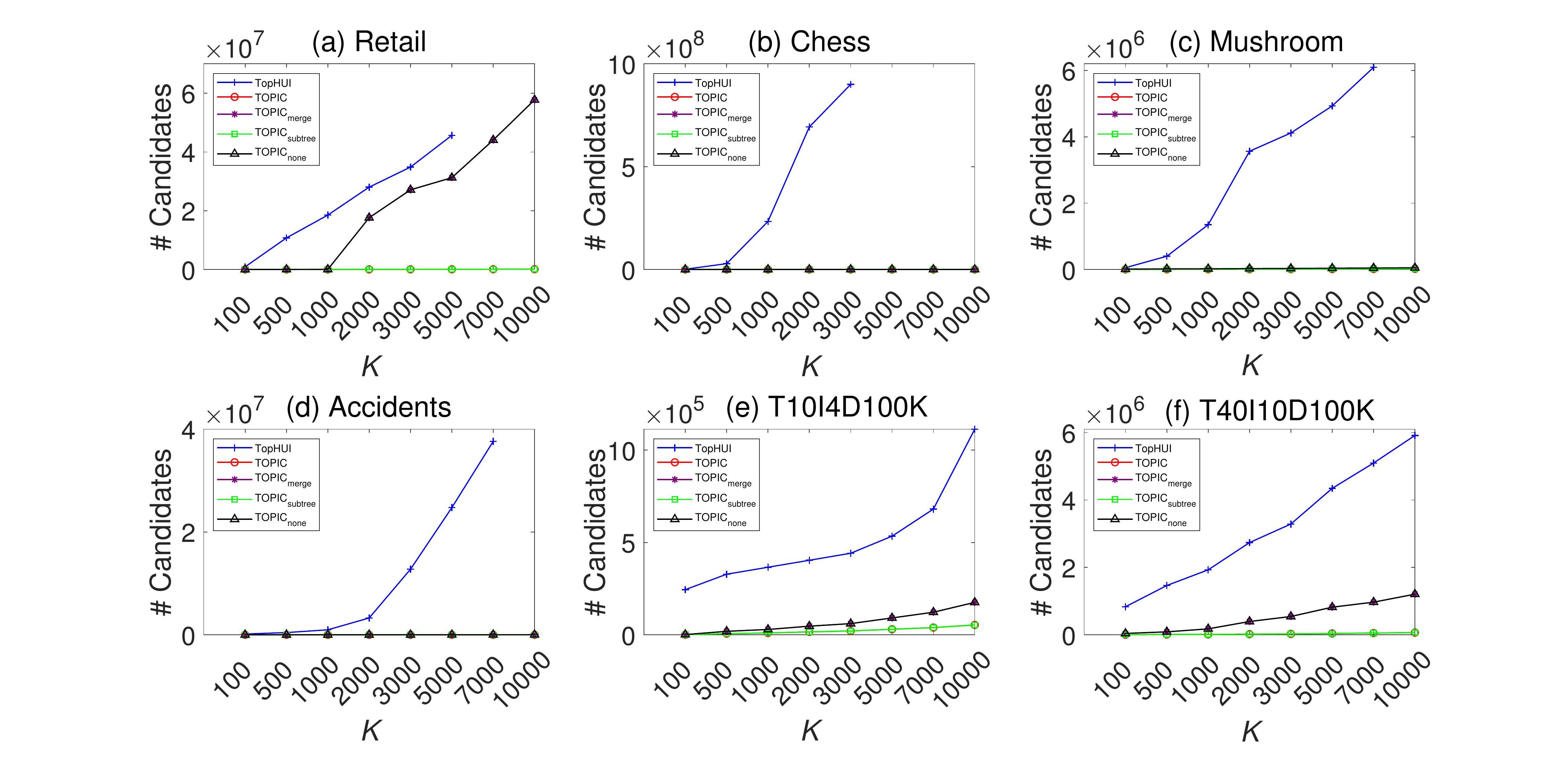}}
	\caption{Candidates under parameter (\textit{K}).}
	\label{fig:Candidates}	
\end{figure*}

\subsection{Experiments on Runtime}
\label{sec:runtime}

First, we evaluated the execution time of the proposed algorithm. Fig. \ref{fig:Runtime} shows the comparison of the runtime of all the algorithms on different datasets with varied $k$ values. In these figures, the TOPIC algorithm performs better than the TopHUI algorithm in many cases. For example, when $k$ is 2000 in the \textit{Mushroom} dataset, TOPIC only requires approximately 2 seconds to complete the mining process, whereas TopHUI requires approximately 6 seconds. On the \textit{Retail} dataset, TopHUI takes too long time to return the result when $k$ is set to 10,000. In all the tested datasets (\textit{Mushroom, Chess, Retail, T10I4D100K, Accidents, T40I10D100K}), the runtime trend between TOPIC and TopHUI becomes increasingly different as the $k$ values increase. In most datasets, TOPIC usually has a narrower fluctuation margin of the exchange rate.

\begin{table*}[!h]
	\caption{Runtime result (seconds)}
	\centering
	
	\label{table:Runtime}
	\setlength{\tabcolsep}{3.8mm}{
		\begin{tabular}{llllllll}
			\toprule
			\textbf{\textit{K}} & \textbf{Algorithm} & \textbf{Retail} & \textbf{Chess} & \textbf{Mushroom} & \textbf{T10I4D100K} & \textbf{Accidents} & \textbf{T40I10D100K}  \\ 
			\midrule
			\multirow{5}*{100}   
			& TopHUI    		 & 4.545  & 12.269  & 1.082 & 4.533 & 168.629 & 37.684 \\
			& TOPIC     		 & 3.894  & 1.115   & 0.953 & 4.197 & 71.484  & 9.371  \\
			& TOPIC$_{merge}$	 & 9.985  & 3.01    & 1.807 & 4.367 & 103.594 & 12.515 \\
			& TOPIC$_{subtree}$  & 2.411  & 0.991   & 0.647 & 1.852 & 63.977  & 8.14   \\
			& TOPIC$_{none}$	 & 4.804  & 2.491   & 1.178 & 2.276 & 93.59   & 10.949 \\ 
			\hline		 
			\multirow{5}*{500}   
			& TopHUI    		 & 17.497 & 134.975 & 2.5   & 6.246 & 479.663 & 45.039 \\
			& TOPIC     		 & 11.818 & 1.614   & 1.016 & 5.237 & 173.129 & 10.429 \\
			& TOPIC$_{merge}$	 & 21.583 & 3.261   & 1.959 & 5.054 & 218.826 & 15.653 \\
			& TOPIC$_{subtree}$  & 6.879  & 1.589   & 1.293 & 3.116 & 158.809 & 8.658  \\
			& TOPIC$_{none}$	 & 12.83  & 3.192   & 1.443 & 2.938 & 203.958 & 13.666 \\
			\hline
			\multirow{5}*{1000}  
			& TopHUI    		 & 24.705 & 46.818  & 3.947 & 6.937 & 821.578 & 46.007 \\
			& TOPIC     		 & 16.557 & 2.117   & 1.274 & 5.017 & 268.598 & 11.508 \\
			& TOPIC$_{merge}$	 & 28.104 & 4.148   & 2.247 & 4.54  & 321.059 & 15.467 \\
			& TOPIC$_{subtree}$  & 9.334  & 2.23    & 1.391 & 2.757 & 250.732 & 10.182 \\
			& TOPIC$_{none}$	 & 14.416 & 3.995   & 1.894 & 2.508 & 290.812 & 13.727 \\
			\hline
			\multirow{5}*{5000}  
			& TopHUI   		 	 & 46.365 & -       & 8.745 & 8.514 &4,437.073& 61.913 \\
			& TOPIC     		 & 18.967 & 4.818   & 2.888 & 3.868 & 767.916 & 16.38  \\
			& TOPIC$_{merge}$	 & 35.193 & 7.506   & 3.447 & 4.111 & 853.639 & 21.797 \\
			& TOPIC$_{subtree}$  & 18.925 & 5.398   & 3.349 & 2.867 & 713.551 & 14.858 \\
			& TOPIC$_{none}$	 & 27.036 & 6.888   & 3.737 & 2.853 & 897.455 & 20.361 \\
			\hline
			\multirow{5}*{10000} 
			& TopHUI    		 & -      & -       & -     & 9.862 & -       & 70.921 \\
			& TOPIC     		 & 23.744 & 7.892   & 4.117 & 4.107 &1,233.659& 20.703 \\
			& TOPIC$_{merge}$	 & 41.208 & 10.512  & 4.85  & 4.521 &1,343.106& 27.303 \\
			& TOPIC$_{subtree}$  & 22.614 & 9.042   & 4.987 & 3.086 &1,146.068& 18.76  \\
			& TOPIC$_{none}$	 & 34.354 & 10.156  & 5.482 & 3.339 &1,253.623& 25.641 \\
			\bottomrule
	\end{tabular}}
\end{table*}

From Table \ref{table:Runtime}, TOPIC performs particularly well on dense and moderately dense datasets. As the parameter $K$ increases, the runtime cost of all the algorithms becomes increasingly higher. Their results in Figure \ref{fig:Runtime} shows TopHUI raises faster, but TOPIC is raising smooth except TOPIC$_{none}$. Generally, TOPIC is approximately one to three orders of magnitude faster than TopHUI. For the \textit{Retail, Chess, Mushroom} and \textit{T10I4D100K} datasets, TOPIC is up to 3, 20, 4, and 2 times faster than the TopHUI algorithm.

\begin{table*}[!h]
	\caption{Memory cost (MB)}
	\centering
	
	\label{table:Memory}
	\setlength{\tabcolsep}{3.8mm}{
		\begin{tabular}{llllllll}
			\toprule
			\textbf{\textit{K}} & \textbf{Algorithm} & \textbf{Retail} & \textbf{Chess} & \textbf{Mushroom} & \textbf{T10I4D100K} & \textbf{Accidents} & \textbf{T40I10D100K}  \\  
			\midrule
			\multirow{5}*{100}   
			& TopHUI   		 	 & 373.95 &1,336.48 & 516.9  & 301.8  & 2112    &1,423.43 \\
			& TOPIC     		 & 69.63  & 66.09   & 66.91  & 99.8   &1,360.61 & 551.7   \\
			& TOPIC$_{merge}$	 & 80.04  & 65.97   & 67.28  & 114.07 &1,373.68 & 599.09  \\
			& TOPIC$_{subtree}$  & 69.64  & 65.95   & 66.97  & 99.12  &1,356.82 & 551.73  \\
			& TOPIC$_{none}$	 & 77.63  & 139.4   & 130.54 & 114.06 &1,380.66 & 599.33  \\ 
			\hline
			\multirow{5}*{500}   
			& TopHUI    		 & 636.49 &1,306.25 &1,328.7 & 356.42 &2,624.92 &1,443.47 \\
			& TOPIC   		 	 & 95.45  & 66.02   & 130.7  & 134.01 &1,396.46 & 551.45  \\
			& TOPIC$_{merge}$	 & 105.38 & 129.57  & 136.55 & 143.39 &1,405.88 &1,065.99 \\
			& TOPIC$_{subtree}$  & 95.33  & 129.55  & 69.68  & 136.48 &1,404.48 & 550.33  \\
			& TOPIC$_{none}$	 & 103.39 & 129.51  & 133.79 & 143.39 &1,390.12 & 845.97  \\ 
			\hline
			\multirow{5}*{1000}  
			& TopHUI    		 & 642.96 &1,333.29 &1,331.07& 416.84 &3,322.79 &1,461.81 \\
			& TOPIC     		 & 107.7  & 128.72  & 130.81 & 143.38 &1,410.31 & 551.74  \\
			& TOPIC$_{merge}$	 & 118.71 & 129.51  & 257.3  & 143.39 &1,449.51 &1,065.49 \\
			& TOPIC$_{subtree}$  & 106.1  & 129.59  & 129.77 & 143.38 &1,406.9  & 551.69  \\
			& TOPIC$_{none}$	 & 117.69 & 129.59  & 257.38 & 143.38 &1,439.16 &1,065.58 \\ 
			\hline
			\multirow{5}*{5000}  
			& TopHUI    		 & 803.66  & -      &1,330.19& 556.73 &3,837.37 &1,497.28 \\
			& TOPIC     		 & 201.84  & 257.42 & 258.2  & 143.38 &1,379.98 &1,065.81 \\
			& TOPIC$_{merge}$	 &1,329.31 & 256.93 & 511.73 & 145.87 &1,381.22 &1,321.86 \\
			& TOPIC$_{subtree}$  & 199.38  & 257.35 & 257.81 & 143.39 &1,380.56 &1,065.74 \\
			& TOPIC$_{none}$	 &1,328.95 & 257.3  & 258.33 & 155.92 &1,354.89 &1,321    \\ 
			\hline
			\multirow{5}*{10000} 
			& TopHUI    		 & -       & -      & -      & 572.66 & -		&1,502.87 \\
			& TOPIC     		 & 275.03  & 511.8  & 512.94 & 181.01 &1,344.19 &1,066.54 \\
			& TOPIC$_{merge}$	 &1,330.7  & 257.98 & 512.7  & 220.63 &1,369.76 &1,322.49 \\
			& TOPIC$_{subtree}$  & 275.03  & 257.21 & 258.64 & 195.88 &1,373.93 &1,066.19 \\
			& TOPIC$_{none}$	 &1,330.12 & 511.83 & 258.61 & 243.4  &1,471.99 &1,322.48 \\ 
			\bottomrule
	\end{tabular}}
\end{table*}

\begin{table*}[!h]
	\caption{Candidates generation}
	\centering
	
	\label{table:Candidate}
	\setlength{\tabcolsep}{3.8mm}{
		\begin{tabular}{llllllll}
			\toprule
			\textbf{\textit{K}} & \textbf{Algorithm} & \textbf{Retail} & \textbf{Chess} & \textbf{Mushroom} & \textbf{T10I4D100K} & \textbf{Accidents} & \textbf{T40I10D100K}  \\  
			\midrule
			\multirow{5}*{100}   
			& TopHUI   		 	 &902,981 &1,095,506&52,751 &244,575 &158,950 &83,360 \\
			& TOPIC     		 &1,105   &7,578    &1,822  & 551 	 &1,438   &4,759  \\
			& TOPIC$_{merge}$	 &2,949   &63,098   &10,686 &1,438   &4,137   &42,021 \\
			& TOPIC$_{subtree}$  &1,105   &7,578    &1,822  & 551    &1,438   &4,759  \\
			& TOPIC$_{none}$	 &2,949   &63,098   &10,686 &1,438   &4,137   &42,021 \\ 
			\hline
			\multirow{5}*{500}   
			& TopHUI    		 &10,747,653&28,358,203&399,129 &328,210&441,707 &1,463,994\\
			& TOPIC     		 &5,044     &11,132    &3,533   &6,543  &2,385   &9,774    \\
			& TOPIC$_{merge}$	 &11,311    &83,155    &16,479  &19,111 &6,614   &88,898   \\
			& TOPIC$_{subtree}$  &5,044     &11,132    &3,533   &6,543  &2,385   &9,774    \\
			& TOPIC$_{none}$	 &11,311    &83,155    &16,479  &19,111 &6,614   &88,898   \\ 
			\hline
			\multirow{5}*{1000}  
			& TopHUI    		 &18,518,220&3,522,998&1,348,989&365,864&981,799 &1,928,124\\
			& TOPIC     		 &9,661     &13,521   &4,832    &9,981  &2,923   &13,118   \\
			& TOPIC$_{merge}$	 &22,454    &96,391   &20,236   &29,096 &8,229   &177,968  \\
			& TOPIC$_{subtree}$  &9,661     &13,521   &4,832    &9,981  &2,923   &13,118   \\
			& TOPIC$_{none}$	 &22,454    &96,391   &20,236   &29,096 &8,299   &177,968  \\ 
			\hline
			\multirow{5}*{5000}  
			& TopHUI    		 &45,661,581& -      &493,057 &533,965 &24,758,053&4,345,977\\
			& TOPIC     		 &60,025    &22,067  &12,313  &30,181  &5,101     &41,525   \\
			& TOPIC$_{merge}$	 &31,247,162&135,472 &37,989  &91,556  &14,428    &820,906  \\
			& TOPIC$_{subtree}$  &60,025    &22,067  &12,313  &30,181  &5,101     &41,525   \\
			& TOPIC$_{none}$	 &31,247,162&135,472 &37,989  &91,556  &14,428    &820,906  \\
			\hline
			\multirow{5}*{10000} 
			& TopHUI    		& -        & -      & -     &1,112,932& -     &5,914,611 \\
			& TOPIC    		 	&93,323    &27,598  &18,146 &53,488   &6,865  &66,951    \\
			& TOPIC$_{merge}$	&57,730,102&158,524 &52,090 &175,800  &18,736 &1,206,662 \\
			& TOPIC$_{subtree}$ &93,323    &27,598  &18,146 &53,488   &6,865  &66,951    \\
			& TOPIC$_{none}$	&57,730,102&158,524 &52,090 &175,800  &18,736 &1,206,662 \\
			\bottomrule
	\end{tabular}}
\end{table*}

The most important reason why TOPIC has an excellent performance in all the datasets is that it utilizes the \textit{RSU} and \textit{RLU} upper bounds, depending on the projected database. It can prune a larger part of the search space compared to the TopHUI algorithm, which uses different strategies. Therefore, the proposed algorithm uses only a few itemsets to find high utility itemsets. It also utilizes a transaction merging technique to replace some transactions (which have identical items) with one transaction, which significantly reduces the cost of dataset scanning.

\subsection{Experiments on Memory Evaluation}
\label{sec:memory}

In this subsection, the memory usage of all the tested algorithms is recorded and compared with the varying $K$ parameter. Fig. \ref{fig:Memory} shows the detailed result, in which TOPIC clearly outperforms TopHUI on all the datasets. For example, in Fig. \ref{fig:Memory}(b), \textit{Chess} dataset reveals that no matter the strategies that TOPIC adopts, TOPHUI uses almost eight times more memory than TOPIC to complete data mining. Moreover, in \textit{Chess} datasets, when $K$ is more than 5000, TopHUI cannot obtain the correct result in regular time (approximately 3 h).

Table \ref{table:Memory} shows more details. In \textit{T10I4D100K}, TOPIC uses 3.0, 2.6, 2.9, 3.8, and 3.2 times less memory than TopHUI when the $k$ parameter is 100, 500, 1000, 5000, and 10,000 respectively. The worst case is that TopHUI cannot obtain the correct results when $K$ is 10,000 in all the tested datasets. It is also interesting that TOPIC has the same performance as TOPIC$_{subtree}$, and TOPIC$_{merge}$ has the same performance as TOPIC$_{none}$. This s because sub-tree pruning strategy plays a vital role. While it iterates the searching space tree, if the \textit{RTWU} value of an itemset is less than the current \textit{minUtil}, sub-tree pruning strategy will remove it and its supersets. Because of the basis of the \textit{RTWU}-based pruning, if an itemset $X$ is not a HUI, all supersets of $X$ would be low-utility itemsets.

Another reason why TOPIC performs quite efficiently is that it adopts a completely different data structure, which does not need to maintain a large amount of information in the memory. It only requires pointers for pseudo-projections to catch the pre-HUIs. However, TopHUI relies on a list structure to store additional information, which is more complex. It also calculates two utility upper bounds in linear time by arrays, which can be repeatedly used to count the upper bounds of each itemset while processing the depth-first search. These new upper bounds help to select extensible itemsets and potential HUIs to ignore unpromising itemsets.

\subsection{Experiments on Candidates Analysis}
\label{sec:candidates}

We also compared the ability of the TOPIC and TopHUI algorithms to prune the search space. Table \ref{table:Candidate} summarizes the results of TopHUI, TOPIC$_{merge}$, TOPIC$_{subtree}$, TOPIC$_{none}$, and TOPIC when $K$ is 100, 500, 1000, 5000, and 10,000, respectively. It can be observed that TOPIC$_{merge}$, TOPIC$_{subtree}$, TOPIC$_{none}$, and TOPIC are more effective than TopHUI when pruning the search space. This is because the TOPIC algorithm adopts two special tight upper bounds (\textit{RSU} and \textit{RLU}). Upper bounds help to remove these low utility itemsets because they are irrelevant. In Table \ref{table:Candidate}, each column shows that TopHUI generates beyond one to three orders of magnitude compared to TOPIC. Fig. \ref{fig:Candidates} shows the rough trend of all outputs of the compared algorithms.

\subsection{Experiments on Scalability Test}
\label{sec:scalability}

\begin{figure}[!h]
	\centering 
	\scalebox{0.88}{\includegraphics[trim={2cm 0cm 1cm 0cm}, clip, width=0.51\textwidth,height=0.33\textwidth]{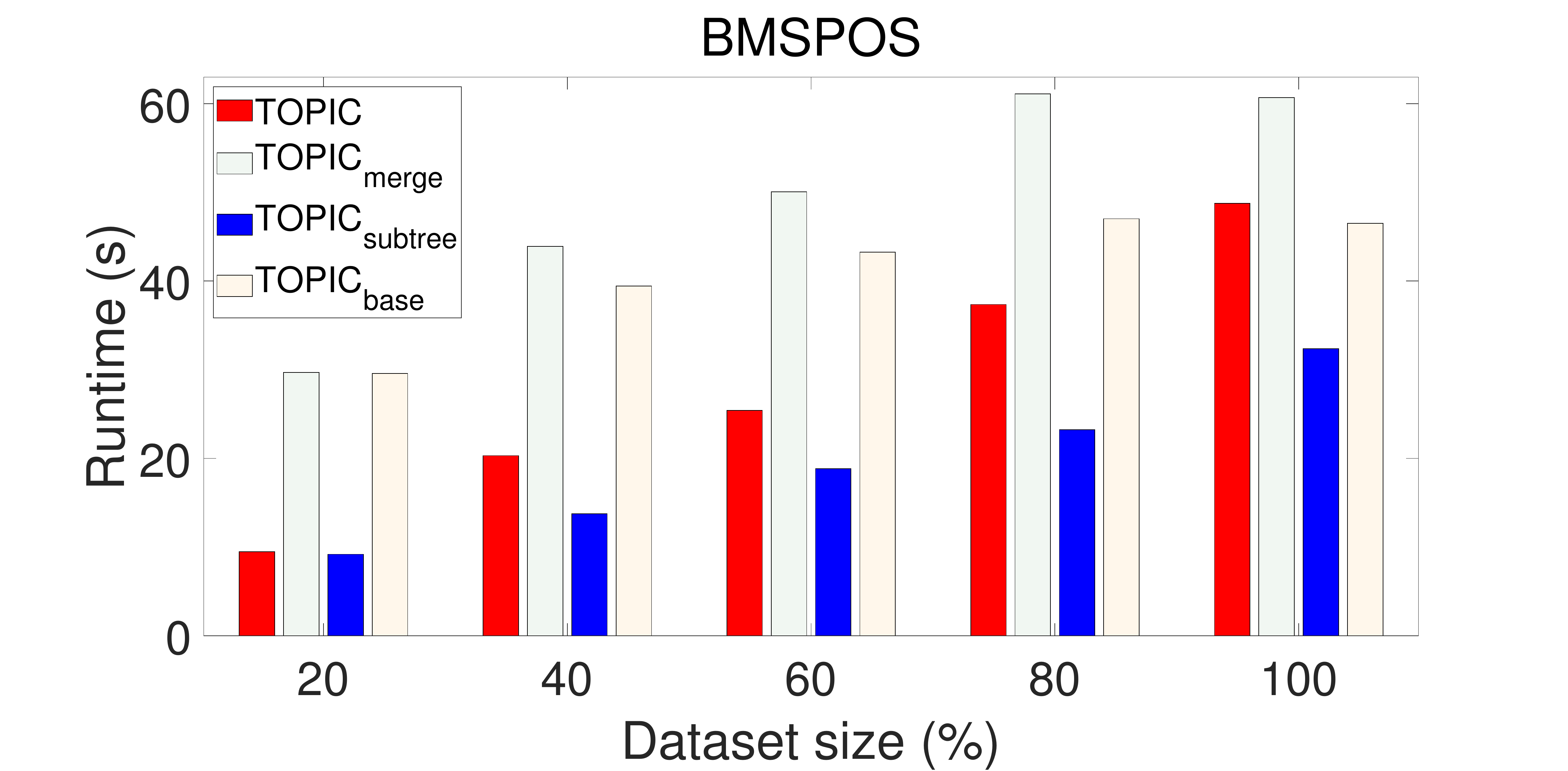}}
	\caption{Runtime scalability of algorithms on \textit{BMSPOS}.}
	\label{fig:Scalability_runtime}	
\end{figure}

\begin{figure}[!h]
	\centering
	\scalebox{0.88}{\includegraphics[trim={1cm 0cm 1cm 0cm}, clip, width=0.52\textwidth,height=0.3\textwidth]{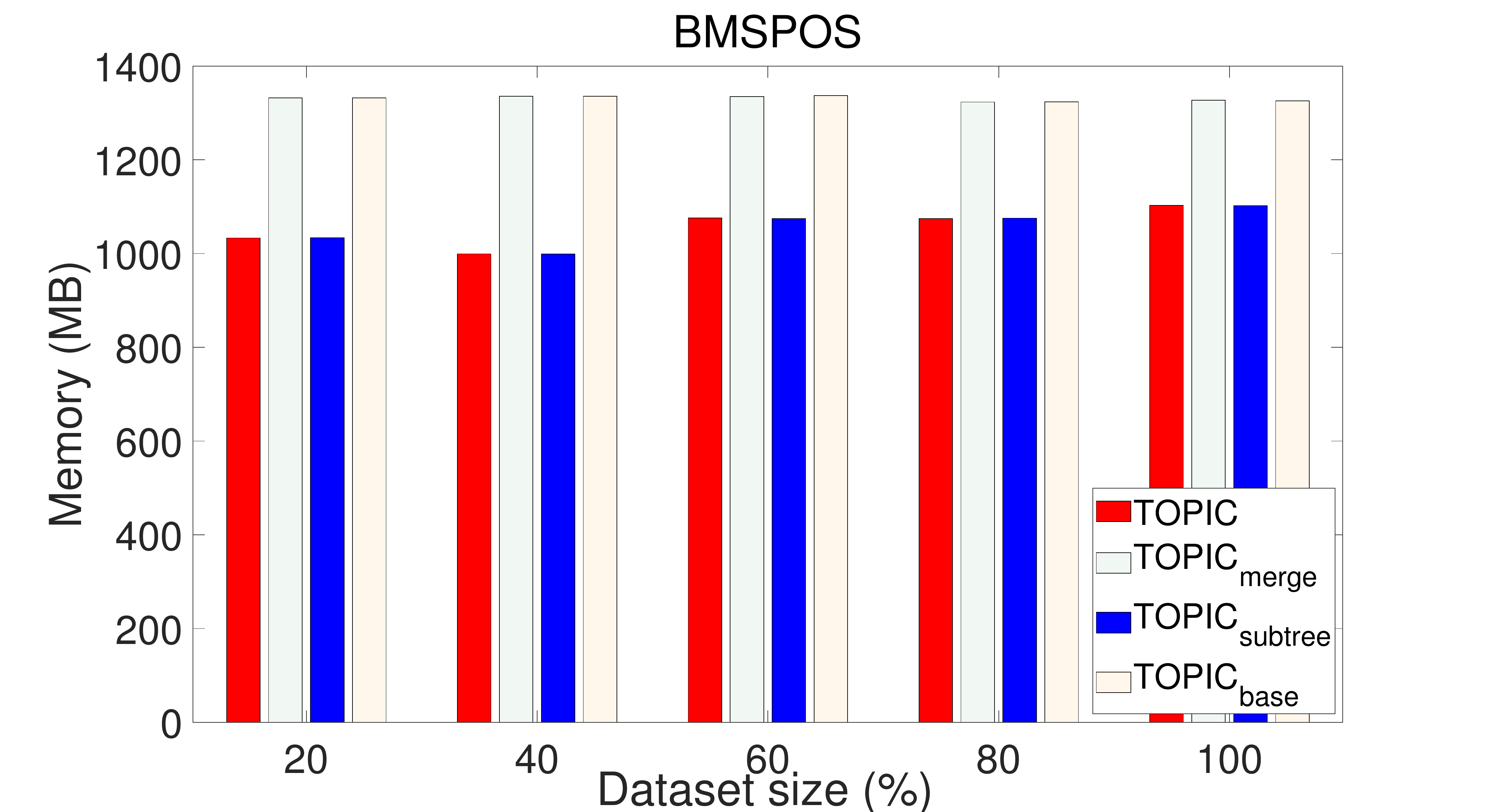}}
	\caption{Memory scalability of algorithms on \textit{BMSPOS}.}
	\label{fig:Scalability_memory}	
\end{figure}

Finally, we tested the scalability of TOPIC. We varied the size of the \textit{BMSPOS} dataset from 20\% (= 103,073 transactions) to 100\% (= 515,366 transactions), and compared the execution time and memory consumption. We set the value of $K$ to 10,000 to check the scalability performance of four variants, such as TOPIC, TOPIC$_{merge}$, TOPIC$_{subtree}$, and TOPIC$_{none}$. Figs. \ref{fig:Scalability_runtime} and \ref{fig:Scalability_memory} separately show that the runtime and memory cost increase linearly with increased dataset size. In particular, both the runtime and memory consumption of TOPIC performed better than others. Thus, TOPIC has suitable scalability for large-scale datasets.

\section{Conclusion and Future Work}
\label{sec:conclusion}

In this work, top-$k$ HUI mining with negative utility was proposed. Our proposed algorithm, TOPIC, adopts two new upper bounds called redefined local utility and redefined sub-tree utility to quickly prune the search space. In addition, we utilized novel utility arrays to efficiently calculate these upper bounds. To reduce the costs of dataset scanning and memory, we adopted dataset projection and transaction merging techniques. Without setting threshold, \textit{minUtil} threshold auto-raising strategy was utilized. Compared with state-of-the-art algorithms, the results show that TOPIC has a significantly improved runtime performance on real and synthetic datasets. Moreover, the memory consumption of TOPIC on all datasets was excellent. 

In the future, we will improve the threshold auto-raising strategy and design more compressed data structures. The proposed idea can also be used in the field of on-shelf utility mining, incremental mining of HUIs, and mining of top-$k$ HUIs from data streams or sequential datasets.

\ifCLASSOPTIONcaptionsoff
  \newpage
\fi


\bibliographystyle{IEEEtran}
\bibliography{topic}

%

%


\vspace{-1.5cm}
\begin{IEEEbiography}[{\includegraphics[width=1in,height=1.25in,clip,keepaspectratio]{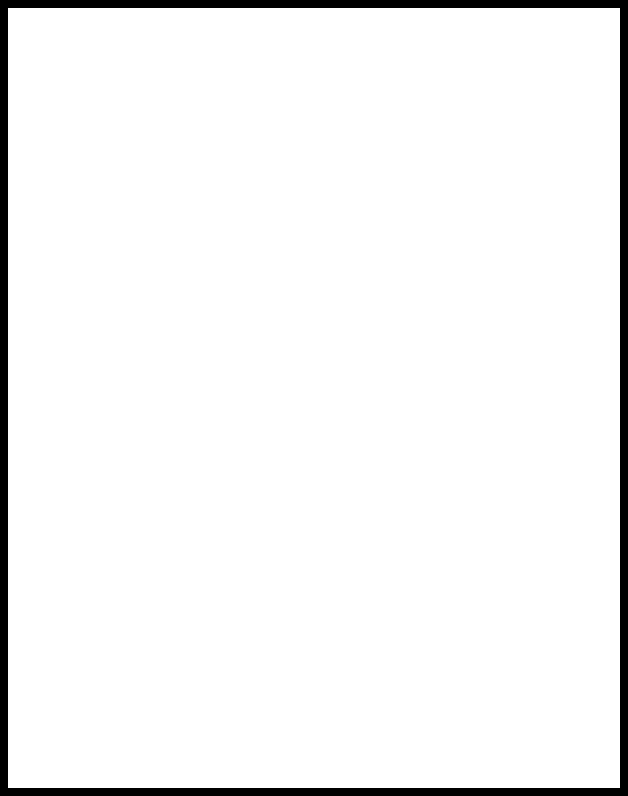}}]{Jiahui Chen (Member, IEEE)}
	received the BS degree from South China Normal University, China in 2009, and MS and PhD degrees from South China University of Technology, China, in 2012 and 2016, respectively. He joined National University of Singapore as a research scientist between form 2017 to 2018. He is currently an associate professor in the School of Computer Sciences, Guangdong University of Technology, China. His research interests mainly focus on public key cryptography, post-quantum cryptography, and information security. 
\end{IEEEbiography}

\vspace{-1.5cm}
\begin{IEEEbiography}[{\includegraphics[width=1in,height=1.25in,clip,keepaspectratio]{newAuthor.png}}]{Shicheng Wan} 
	received the B.S. degree in Gannan Normal University, Jiangxi, China in 2020. He is currently a postgraduate in the School of Computer Sciences, Guangdong University of Technology, China. His research interests include data mining, utility mining, and big data. 
\end{IEEEbiography}

\vspace{-1.5cm}
\begin{IEEEbiography}[{\includegraphics[width=1in,height=1.25in,clip,keepaspectratio]{newAuthor.png}}]{Wensheng Gan  (Member, IEEE)} 
	received the B.S. degree in Computer Science from South China Normal University, China in 2013. He received the Ph.D. in Computer Science and Technology, Harbin Institute of Technology (Shenzhen), China in 2019. He was a joint Ph.D. student with the University of Illinois at Chicago, Chicago, USA, from 2017 to 2019. He is currently an Association Professor with the College of Cyber Security, Jinan University, Guangzhou, China.  His research interests include data mining, utility computing, and big data analytics. He has published more than 80 research papers in peer-reviewed journals (i.e., IEEE TKDE, IEEE TCYB, ACM TKDD, ACM TOIT, ACM TMIS) and international conferences. He is an Associate Editor of \textit{Journal of Internet Technology}. 
\end{IEEEbiography}

\vspace{-1.5cm}
\begin{IEEEbiography}[{\includegraphics[width=1in,height=1.25in,clip,keepaspectratio]{newAuthor.png}}]{Guoting Chen}  
	 is currently a full professor with School of Science, Harbin Institute of Technology, Shenzhen. He received B.S., M.S. and Ph.D. degrees in Mathematics from Wuhan University, China in 1982, from Wuhan University, China in 1985, and from University de Grenoble 1, France in 1990, respectively. His research interests include Mathematics, differential equations, and data science. He has published 30 peer-reviewed research papers. 
\end{IEEEbiography}

\vspace{-1.5cm}
\begin{IEEEbiography}[{\includegraphics[width=1in,height=1.25in,clip,keepaspectratio]{newAuthor.png}}]{Hamido Fujita  (Senior Member, IEEE)}  
	is currently a Professor with Iwate Prefectural University, Takizawa, Japan, as Director of Intelligent Software Systems. He received Doctor Honoris Causa from Óbuda University, Budapest, Hungary, in 2013 and received Doctor Honoris Causa from Timisoara Technical University, Timisoara, Romania, in 2018, and a title of Honorary Professor from Óbuda University, in 2011. He is the Emeritus Editor-in-Chief for Knowledge-Based Systems, and currently Editor-in-Chief of Applied Intelligence (Springer), He is the Vice President of International Society of Applied Intelligence. He headed a number of projects including intelligent HCI, a project related to mental cloning for healthcare systems as an intelligent user interface between human-users and computers, and SCOPE project on virtual doctor systems for medical applications. He has published more 400 highly cited Papers.
\end{IEEEbiography}

\end{document}